\documentclass[10pt,journal]{IEEEtran}
\usepackage{graphics,color,algorithmic,algorithm}

\usepackage{nonfloat}
\usepackage{graphicx}
\usepackage{amssymb,amsmath,amsfonts,amssymb}
\usepackage{amsmath,amssymb,mathrsfs,bm,url,times,multirow}
\usepackage{subfigure,wrapfig}
\newcommand{\E}{{\sf E}}

\begin{document}

\title{Characterizing Honeypot-Captured Cyber Attacks: Statistical Framework and Case Study}

\author{Zhenxin Zhan, Maochao Xu, and Shouhuai Xu
\thanks{Copyright (c) 2013 IEEE. Personal use of this material is permitted. However, permission to use this material for any other purposes 
must be obtained from the IEEE by sending a request to pubs-permissions@ieee.org}
\thanks{Zhenxin Zhan and Shouhuai Xu are with the Department of Computer Science,
University of Texas at San Antonio, San Antonio, TX 78249. Emails: {\tt jankins.ics@gmail.com} (Zhenxin Zhan), {\tt shxu@cs.utsa.edu} (Shouhuai Xu; corresponding author)}
\thanks{Maochao Xu is with the Department of Mathematics, Illinois State University, Normal, IL 61790.
Email: {\tt mxu2@ilstu.edu}}}

\maketitle

\begin{abstract}
Rigorously characterizing the statistical properties of cyber attacks is an important problem.
In this paper, we propose the {\em first} statistical framework for rigorously analyzing
honeypot-captured cyber attack data.
The framework is built on the novel concept of {\em stochastic cyber attack process}, a new kind of mathematical objects
for describing cyber attacks.
To demonstrate use of the framework, we apply it to analyze a low-interaction honeypot dataset, while noting that the framework
can be equally applied to analyze high-interaction honeypot data that contains richer information about the attacks.
The case study finds, for the first time, that Long-Range Dependence (LRD) is exhibited by honeypot-captured cyber attacks.
The case study confirms that by exploiting the statistical properties (LRD in this case),
it is feasible to predict cyber attacks (at least in terms of attack rate) with good accuracy.
This kind of prediction capability would provide
sufficient early-warning time for defenders to adjust their defense configurations or resource allocations.
The idea of ``gray-box" (rather than ``black-box") prediction  is central to the utility of the statistical framework,
and represents a significant step towards ultimately understanding (the degree of) the {\em predictability} of cyber attacks.
\end{abstract}


\IEEEpeerreviewmaketitle

\section{Introduction}

Characterizing statistical properties of cyber attacks not only can deepen our understanding of cyber threats
but also can lead to implications for effective cyber defense.
Honeypot is an important tool for collecting cyber attack data,
which can be seen as a ``birthmark" of the cyber threat landscape as observed from a certain IP address space.
Studying this kind of data allows us to extract useful information about, and even predict, cyber attacks.
Despite the popularity of honeypots, there is no systematic framework for
rigorously analyzing the statistical properties of honeypot-captured cyber attack data.
This may be attributed to that a systematic framework would require both a nice abstraction of cyber attacks and fairly advanced statistical techniques.

In this paper, we make three contributions.
First, we propose, to our knowledge, the first statistical framework for systematically analyzing and exploiting honeypot-captured cyber attack data.
The framework is centered on the concept we call
{\em stochastic cyber attack process}, which is a new kind of mathematical objects that can naturally model cyber attacks.
This concept can be instantiated at multiple resolutions, such as: network-level (i.e., considering all attacks against a network as a whole),
victim-level (i.e., considering all attacks against a computer or IP address as a whole),
port-level (i.e., the defender cares most about the attacks against certain ports or services).
This concept catalyzes the following fundamental questions:
(i) What statistical properties do stochastic cyber attack processes exhibit (e.g., are they Poisson)?
(ii) What are the implications of these properties and, in particular, can we exploit them to predict the incoming attacks
(prediction capability is the core utility of the framework)?
(iii) What caused these properties?
Thus, the present paper formulates a way of thinking for rigorously analyzing honeypot data.

Second, we demonstrate use of the framework by applying it to analyze a
dataset, which is collected by a {\em low-interaction} honeypot of 166 IP addresses for five periods of time (220 days cumulative).
Findings of the case study include:
(i) Stochastic cyber attack processes are not Poisson, but instead can exhibit Long-Range Dependence (LRD) --- a property that is not known to be exhibited by honeypot data
until now. This finding has profound implications for modeling cyber attacks.
(ii) LRD can be exploited to predict the incoming attacks at least in terms of attack rate (i.e., number of attacks per time unit).
This is especially true for network-level stochastic cyber attack processes.
This shows the power of ``gray-box" prediction, where the prediction models accommodate the LRD property
(or other statistical properties that are identified).
(iii) Although we cannot precisely pin down the cause of the LRD exhibited by honeypot data,
we manage to rule out two possible causes.
We find that the cause of LRD exhibited by cyber attacks might be different from
the cause of LRD exhibited by benign traffic (see Section \ref{sec:step-5}).

Third, the framework can be equally applied to analyze both {\em low-interaction} and {\em high-interaction} honeypot data,
while the latter contains richer information about attacks and allows even finer-resolution analysis.
Thus, we plan to make our statistical framework software code publicly available
so that other researchers or even practitioners, who have (for example) high-interaction honeypot data that often cannot be shared with
third parties, can analyze their data without learning the advanced statistic skills.

The paper is organized as follows.
Section \ref{sec:preliminaries} briefly reviews some statistical preliminaries including prediction accuracy measures,
while some detailed statistical techniques are deferred to the Appendix.
Section \ref{sec:concept-and-framework} describes the framework.
Section \ref{sec:case-study} discusses the case study and its limitations.
Section \ref{sec:total-discussion} discusses the limitation of the case study (which is imposed by the
specific dataset) and the usefulness of the framework in a broader context.
Section \ref{sec:related-work} discusses related prior work.
Section \ref{sec:conclusion} concludes the paper with future research directions.

\section{Statistical Preliminaries}
\label{sec:preliminaries}

\subsection{Long-Range Dependence (LRD)}

A stationary time sequence $\{X_t: t\ge 0\}$, which instantiates a stochastic cyber attack process $\{{\bf X}_t:t\geq 0\}$,
is said to possess LRD \cite{SamorodnitskyBook06,WTLW95} if its autocorrelation function
\begin{equation}\label{lrd}
\rho(h)={\rm Cor}(X_t,X_{t+h})\sim h^{-\beta} L(h),\quad h\to\infty,
\end{equation}
for $0<\beta<1$, where $h$ is called ``lag", $L(\cdot)$ is a slowly varying function
meaning that $\lim_{x\rightarrow \infty} \frac{L(ix)}{L(x)}=1$ for all $i>0$.
Intuitively, LRD says that a stochastic process exhibits persistent correlations,
namely that the rate of autocorrelation decays slowly (i.e., slower than an exponential decay).
Quantitatively speaking, the degree of LRD is expressed by Hurst parameter (H), which is related to the parameter $\beta$  in Eq. \eqref{lrd} as
$\beta=2-2H$  \cite{Beran94}. This means that for LRD, we have $1/2<H<1$ and the degree of LRD increases as $H\rightarrow 1$.
In the Appendix, we briefly review six popular Hurst-estimation methods that are used in this paper.

Since $1/2<H<1$ is necessary but not sufficient for LRD, we need to eliminate the so-called ``spurious LRD" as we focus on the LRD property in this paper.
Spurious LRD can be caused by non-stationarity \cite{Mikosch2004}, or more specifically caused
by (i) short-range dependent time series with change points in the mean or (ii)
slowly varying trends with random noise \cite{Qu2011,Shao2011}.
We eliminate spurious LRD processes by testing the null hypothesis (denoted by $H_0$) that a given time series is a stationary LRD process against the alternative hypothesis
(denoted by $H_a$) that it is affected by change points or a smoothly varying trend \cite{Qu2011}.
One test is for $t\geq 0$:
\begin{eqnarray*}
\begin{array}{ll}
H_0:  & X_t ~\text{is stationary with LRD} \\
H_a:  & X_t=Z_t+\mu_t  ~\text{with}~ \mu_t=\mu_{t-1}+\psi_t\eta_t
\end{array}
\end{eqnarray*}
where $Z_t$ is a stationary short-memory process \cite{CC2008}, $\mu_0=0$,
$\psi_{t}$ is a Bernoulli random variable,
and $\eta_t$ is a white (i.e., Gaussian) noise process.
The other alternative is:
$$\mbox{$H_a$:  $X_t=Z_t+\ell(t/n)$},$$
where $Z_t$ is as in the previous test,
$\ell(\cdot)\in [0,1]$ is a Lipschitz continuous function \cite{Qu2011},
and $n$ is the sample size.

\subsection{Two Statistical Models for Predicting Incoming Attacks}

We call a model {\em LRD-less} if it cannot accommodate LRD and {\em LRD-aware} if it can accommodate LRD.
Let $\epsilon_t$ be independent and identical normal random variables with mean $0$ and variance $\sigma^2_\epsilon$.
We consider two popular models.
\begin{itemize}
\item LRD-less model ARMA$(p,q)$: This is the autoregressive moving average process of orders $p$ and $q$ with
    $$Y_t=\sum_{i=1}^p \phi_i Y_{t-i}+\epsilon_t+\sum_{j=1}^q \theta_j \epsilon_{t-j}.$$
It is one of the most popular models in time series \cite{CC2008}.

\item LRD-aware model FARIMA$(p,d,q)$:
This is the well-known Fractional ARIMA model where $0<d<1/2$ and $H=d+1/2$ \cite{WTLW95,Beran94,AV98}.
Specifically, a stationary process $X_t$ is called FARIMA$(p,d,q)$ if
$$\phi(B)(1-B)^d X_t=\psi(B) \epsilon_t,$$
for some $-1/2<d<1/2$, where
$$\phi(x)=1-\sum_{j=1}^p \phi_j x^j, ~~~~~~ \psi(x)=1+\sum_{j=1}^q \psi_j x^j,$$
$B$ is the back shift operator defined by $B X_t=X_{t-1}$, $B^2 X_t=X_{t-2}$, and so on.
\end{itemize}

\subsection{Measures of Prediction Accuracy}
\label{sec:measures}

Suppose $X_m,X_{m+1},\ldots,X_z$ are observed data (e.g., the attack rate $X_t$ for $m \leq t \leq z$),
and $Y_m,Y_{m+1},\ldots,Y_z$ are the predicted data.
We can define prediction error $e_{t}=X_{t}-Y_{t}$ for $m \leq t \leq z$.
Recall the popular statistic PMAD (Percent Mean Absolute Deviation):
$${\sf PMAD}=\frac{\sum_{t=m}^z |e_t|}{\sum_{t=m}^z X_t},$$
which can be seen as the overall prediction error. We also define
a variant of it, called {\em underestimation error}, which considers only the underestimations as follows:
$${\sf PMAD}'=\frac{\sum_{t=m}^z e_t}{\sum_{t=m}^z X_t} ~~for~e_t>0~and~corresponding~X_t.$$
Underestimation error is useful especially when the defender is willing to over-provision some defense resources
and is more concerned with the attacks that can be overlooked because of insufficient provisioning of defense resources
(e.g., when the attack rate is high and beyond the processing capacity of the defender's provisioned defense resources,
the defender may have to skip examining the traffic in order not to disrupt the services in question).
It is also convenient to use the following {\em overall accuracy} measure ({\sf OA} for short)
and {\em underestimation accuracy} measure ({\sf UA} for short):
$${\sf OA}=1-{\sf PMAD}, ~~~~{\sf UA}=1-{\sf PMAD}'.$$

\section{The Statistical Framework}
\label{sec:concept-and-framework}

\subsection{The Concept of Stochastic Cyber Attack Processes}

Concept at the right level of abstraction is often important.
For describing and modeling cyber attacks, {\em stochastic cyber attack processes} (often called {\em attack processes} for short in the
rest of paper) are a natural abstraction because cyber attack events in principle formulate Point Processes \cite{DaleyBook2003}.
Formally, a stochastic cyber attack process is described as $\{{\bf X}_t: t\geq 0\}$, where ${\bf X}_t$ is the random variable
(e.g., attack rate) at time $t$. Rigorously characterizing the {\em mathematical/probabilistic} properties of stochastic cyber attack processes is
an important problem for theoretical cyber security research, and may not be possible before we have good understandings
about their {\em statistical} properties --- the present paper is one such effort.

Stochastic cyber attack processes can be instantiated at multiple resolutions.
For example, {\em network-level} attack processes accommodate cyber attacks against networks of interest;
{\em victim-level} attack processes accommodate cyber attacks against individual computers or IP addresses;
{\em port-level} attack processes accommodate cyber attacks against individual ports.
The distinction of model resolution is important because a high-level (i.e., low-resolution) attack
process may be seen as the superposition of multiple low-level (i.e., high-resolution) attack processes, which may
help explain the cause, or rule out some candidate causes, of a property exhibited by the high-level process
(see Step 5 in Section \ref{sec:analysis-framework} below for general description and Section \ref{sec:step-5} for case study).

\begin{figure}[!htbp]
\centering
\subfigure[Illustration of victim-level stochastic cyber attack processes with respect to individual victim IP addresses,
where dots represent attack events and (for example) attacks against victim IP 1 arrive at time $t_1,\ldots,t_9$.]
{\includegraphics[width=.48\textwidth]{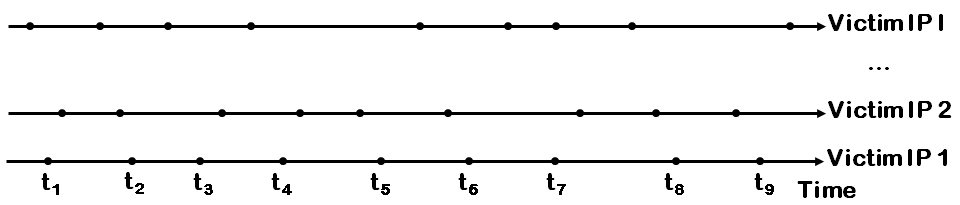}\label{fig:coars-grained-attack}}
\subfigure[Elaboration of a victim-level attack process with respect to victim IP 1.]
{\includegraphics[width=.48\textwidth]{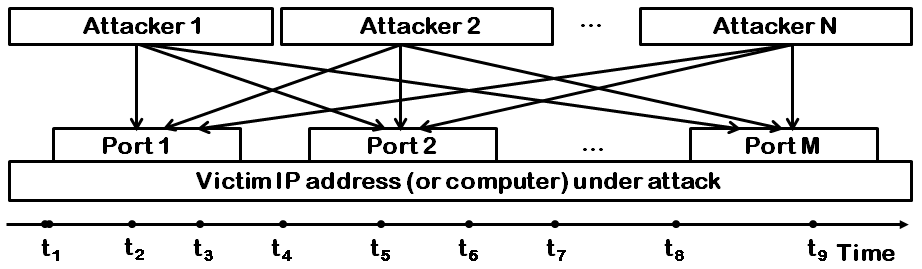}\label{fig:intermediate-grained-attack}}
\caption{Illustration of victim-level stochastic cyber attack processes}
\label{fig:illustration-of-attacks}
\end{figure}

Figure \ref{fig:coars-grained-attack} illustrates the attacks against individual victim IP addresses,
where dots on the same time axis formulate a victim-level attack process.
Figure \ref{fig:intermediate-grained-attack} further shows that a victim is attacked by $N$ attackers (or attacking computers) at some ports
and the attacks arrive at time $t_1,\ldots,t_9$.

\subsection{The Framework}
\label{sec:analysis-framework}

The framework is presented as a 5-step procedure. Step 1 (data pre-processing) is presented for completeness because the data may be collected by software or hardware.
Step 2 (basic statistical analysis) serves the purpose of providing hints for Step 3 (advanced statistical analysis for identifying
statistical properties of attack processes),
which in turn serves as the base for Step 4 (``gray-box" prediction) and Step 5 (exploring the cause of the
newly identified statistical properties).

\paragraph*{\bf Step 1: Data pre-processing}

It is now a common practice to treat honeypot-captured data as attacks because
there are no legitimate services and the honeypot computers passively wait for incoming events.
Honeypot-captured cyber attack data is often organized according to the honeypot IP addresses.
Pre-processing mainly deals with two issues.
First, we may need to differentiate the attack traffic corresponding to the {\em production ports} that are associated to some honeypot programs/services,
and the attack traffic corresponding to the {\em non-production ports} that are not associated to any services.

Second, in order to analyze statistical properties exhibited by honeypot-captured cyber attack data,
we advocate using flows,
rather than IP packets, to represent attacks because of the following.
(i) For low-interaction honeypots data, attack payload is often missing and information about attacks is often captured from the perspective of communication behaviors.
This suggests that flow is appropriate for analyzing honeypot-captured cyber attack data.
(ii) Flow-based intrusion detection is complementary to the traditional packet-based intrusion detection.
For example, flow-based abstraction can be used to detect attacks such as DoS (denial-of-service), scan, worm \cite{flow.worm,DacierRAID08}.
(iii) Flow-based abstraction can deal with encrypted attack payload \cite{flow.ids}, which cannot be dealt with by packet-level analysis.

The concept of flow accommodates both TCP and UDP.
There are COTS devices that can readily extract flows.
However, when honeypot data is collected by software in the format of {\tt pcap} data,
we need to parse it and re-assemble into flows.
Since flow assembly is a standard technique, in what follows we only briefly review the assembly of TCP flows.
A TCP flow is uniquely identified from honeypot-collected raw {\tt pcap}
data via the attacker's IP address, the port used by the
attacker, the victim IP address in the honeypot, and the port that is under attack.
An unfinished TCP handshake can also be treated as a flow (attack) because an unsuccessful handshake can be caused by events
such as: the port in question is busy (i.e., the connection is dropped).
For flows that do not end with the FIN flag (which would indicate safe termination of TCP connection)
or the RST flag (which would indicate unnatural termination of  TCP
connection), we need to choose two parameters in the pre-processing.
One parameter is the {\em flow timeout time}, meaning that a flow is considered expired when
no packet of the flow is received during a time window. For example, 60 seconds would be reasonable for low-interaction honeypots
that provide limited interactions \cite{honeypot.pca}, but a longer time may be needed for high-interaction honeypots.
The other parameter is the {\em flow lifetime}, meaning that a flow is considered expired when a flow
lives longer than a pre-determined lifetime, which can be set as 300 seconds for low-interaction
honeypots \cite{honeypot.pca} but a longer time may be needed for high-interaction honeypots.

\paragraph*{\bf Step 2: Basic statistical analysis}

The basic statistics of cyber attack data can offer hints for advanced statistical analysis.
For stochastic cyber attack processes, the primary statistic is the {\em attack rate}, which describes the number of attacks that arrive at unit time (e.g., minute or hour or day).
Note that attack rate can be instantiated at various resolutions of attack processes, such as:
network-level attack rate, victim-level attack rate and port-level attack rate.
The secondary statistic is the {\em attack inter-arrival time},
which describes the time intervals between two consecutive attack events.
By investigating the $min$, $mean$, $median$, $variance$ and $max$ of these statistics,
we can identify outliers and obtain hints about the properties of the attack processes.
For example, if the attack events are bursty, an attack process may not be Poisson, which can serve as a hint
for further advanced statistical analysis.

\paragraph*{\bf Step 3: Advanced statistical analysis: Identifying statistical properties of attack processes}

This step is to identify statistical properties of attack processes at resolutions of interest.
A particular question that should be asked is: Are the attack processes Poisson? Recall that the Poisson process counts the number of
events that occur during time intervals, where ``events" in the context of this paper are the attacks observed by honeypots.
If not Poisson, what properties do they exhibit?
It would be ideal that the attack processes are Poisson
because we can easily characterize Poisson processes with very few parameters,
and because there are many mature methods and techniques for analyzing them.
For example, we can use the property --- the superposition of
Poisson processes is still a Poisson process \cite{Em1997} --- to simplify problems
when we consider attack processes at multiple resolutions/levels.
In many cases, attack processes may not be Poisson.
For characterizing such processes, we need to use advanced statistical methods, such as
Markov process, L\'{e}vy process, and time-series methods \cite{DaleyBook2003,PR2002}.
This step is crucial because identifying advanced statistical properties can pave the way for answering the next questions.
This step can be quite involved in terms of statistical skills when the attack processes are not Poisson.

\paragraph*{\bf Step 4: Exploiting the statistical properties}

This step addresses the following question: How can we exploit the statistical properties of stochastic cyber attack processes to do useful things?
One exploitation is to conduct ``gray-box" prediction of the incoming attacks, at least in terms of attack rate at the appropriate resolution (which in turn
depends on the data is collected by low-interaction or high-interaction honeypot).
By ``gray-box" prediction we mean that if an attack process exhibits a certain property that is identified in Step 3 (e.g., Long-Range Dependence \cite{SamorodnitskyBook06,WTLW95}
or Short-Range Dependence \cite{PR2002,Beran94}),
the prediction model should accommodate the property as well.
Algorithm \ref{prediction-algorithm} describes a general ``gray-box" prediction algorithm,
where $\{X_1,\ldots,X_t\}$ is the sequence of attack rates observed at time $1,\ldots,t$, and
${\sf h}$ is the number of steps (e.g., hours) we want to predict ahead of time.
In addition to ``gray-box" prediction, Algorithm \ref{prediction-algorithm} is novel also because it selects the best model (from a family of models) at each prediction step,
which is important because there may be no single model that can fit the observed data well at all steps.

\begin{algorithm}[H]
\caption{Prediction Algorithm}
\label{prediction-algorithm}
INPUT: observed attack rates $\{X_1,\ldots,X_t\}$, ${\sf h}$ (steps ahead) \\
OUTPUT: prediction results $Y_{t+{\sf h}},Y_{t+{\sf h}+1},\ldots$ 

\begin{algorithmic}[1]
\REPEAT
\STATE{Fit $\{X_1,\ldots,X_t\}$ to obtain the best model ${\sf M}_t$ from a family of models that accommodate
the newly identified statistical properties (i.e., ``gray-box" prediction) with respect to an appropriate model selection criterion (e.g., Akaike information criterion (AIC) \cite{CC2008})}
\STATE {Use ${\sf M}_t$ to predict $Y_{t+{\sf h}}$, the number of attacks that will arrive during the $(t+{\sf h})$th step}
\STATE {$X_{t+1}\gets$ newly observed attack rate at time $t+1$}
\STATE{ $t\gets t+1$}~~\COMMENT{observing more data as $t$ evolves}
\UNTIL{no need to predict further}
\end{algorithmic}
\end{algorithm}

We note that although the prediction is geared toward honeypot-oriented traffic, it can be useful for defending production networks as well.
This is because when honeypot-captured attacks are increasing (or decreasing), the attack rate with respect to production networks
might also be increasing (or decreasing) as long as the honeypots are approximately uniformly deployed in the IP address space in question.
This can be achieved by blending honeypot IP addresses into production IP addresses.
Since being able to predict incoming attacks (especially hours ahead of time) is always appealing, this would give incentives to deploy honeypots as such.
As a result, it is possible to characterize the relation between the attack traffic into the honeypot IP addresses and the attack traffic into the production IP addresses.

\begin{figure}[!htbp]
\centering
\subfigure[Decomposition of a victim-level attack process into multiple port-level attack processes,
where the attack process corresponding to Port 1 describes the attacks that arrive at time $t_2$ and $t_5$,
the attack process corresponding to Port 2 describes the attacks that arrive at time $t_1$, $t_6$ and $t_9$, etc.]{
\includegraphics[width=.48\textwidth]{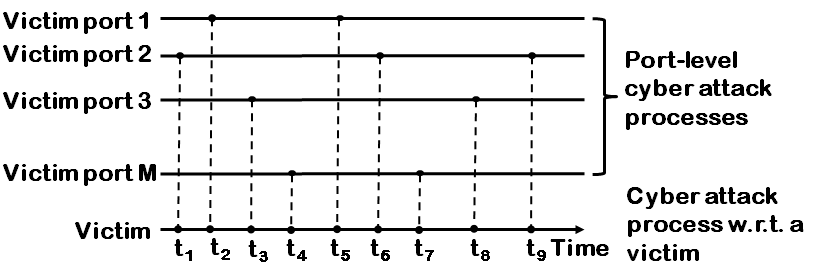}
\label{fig:decomposition-method-port-centric}}
\subfigure[Attacker-level attack process can be derived from victim-level attack process by ignoring the subsequent attacks launched by the same attacker.
In this example, the attacker-level attack process corresponding to the victim describes the attacks that arrive at time $t_1,t_2,t_3,t_4$.]{
\includegraphics[width=.48\textwidth]{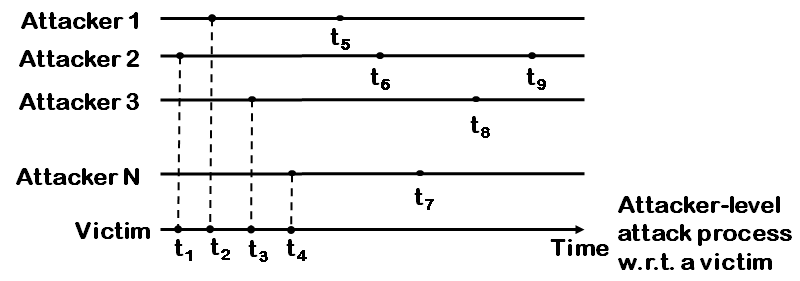}
\label{fig:decomposition-method-attacker-centric}}
\caption{Two approaches to exploring causes of statistical properties}
\end{figure}

\paragraph*{\bf Step 5: Exploring cause of the statistical properties}

This step aims to address the following question: What caused the statistical properties exhibited by stochastic cyber attack processes?
This question is interesting because it reflects a kind of ``natural" phenomenon in cyberspace.
It would be ideal that one can mathematically prove the cause of a property.
This type of ``theoretical proof" approach is often difficult, as witnessed by the outcome of
the past two decades of effort at studying the long-range dependence exhibited by benign Internet traffic (see Section \ref{sec:related-work}).
Therefore, we advocate the ``experimental" approach, which includes the following two specific methods.
The first method is to study the decomposed lower-level (i.e., higher-resolution) stochastic cyber attack processes.
For example, in order to investigate whether or not a certain property is caused by another certain property
of the low-level (i.e., high-resolution) processes,
we can decompose a victim-level attack process
into port-level attack processes that correspond to the individual ports of the victim.
This is illustrated in Figure \ref{fig:decomposition-method-port-centric},
where the victim-level attack process is decomposed into $M$ port-level attack processes.

The second method is to investigate whether or not a certain property is caused by the intense (consecutive) attacks that are launched by individual attackers.
For this purpose, we can consider the attacks against each victim that are launched by {\em distinct} attackers.
As illustrated in Figure \ref{fig:decomposition-method-attacker-centric},
even though an attacker launched multiple consecutive attacks against a victim,
we only need to consider the first attack.
If the attacker-level attack processes do {\em not} exhibit the property that is exhibited by the victim-level attack processes,
we can conclude that the property is probably caused by the intensity of the attacks that are launched by individual attackers.

\begin{figure*}[hbtp]
\subfigure[Period I]{\includegraphics[width=.19\textwidth]{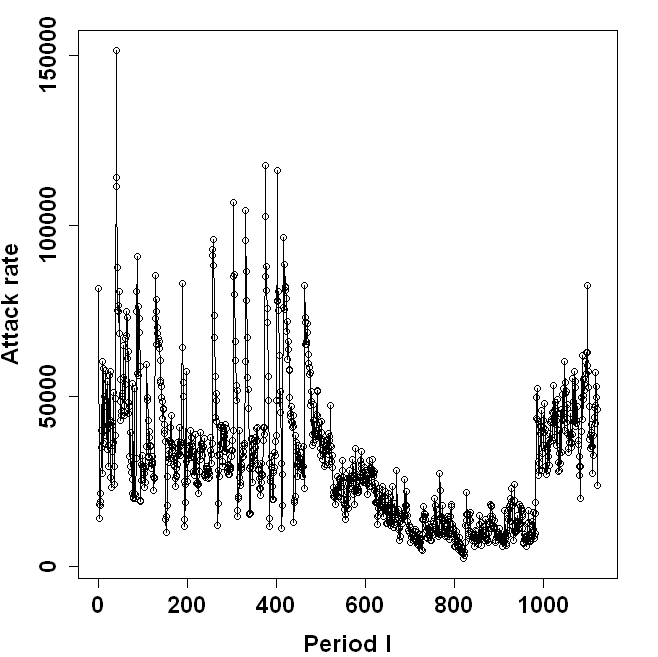}\label{ts-period-I}}
\subfigure[Period II]{\includegraphics[width=.19\textwidth]{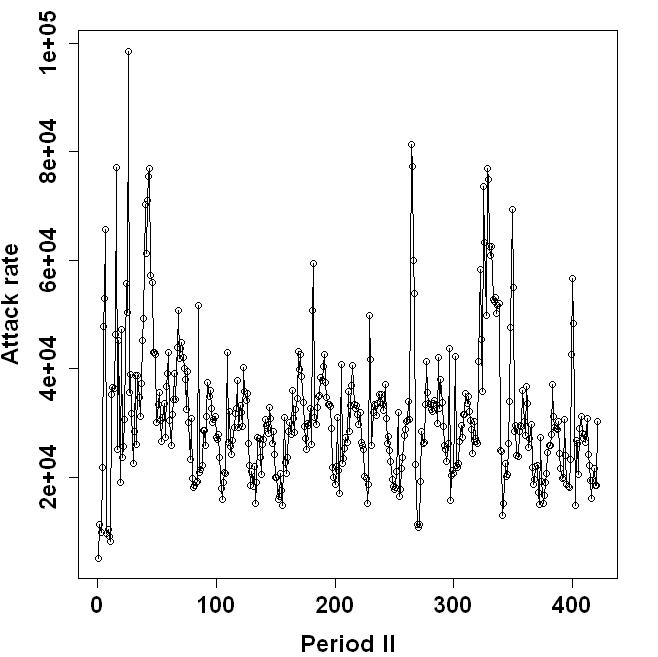} \label{ts-period-II}}
\subfigure[Period III]{\includegraphics[width=.19\textwidth]{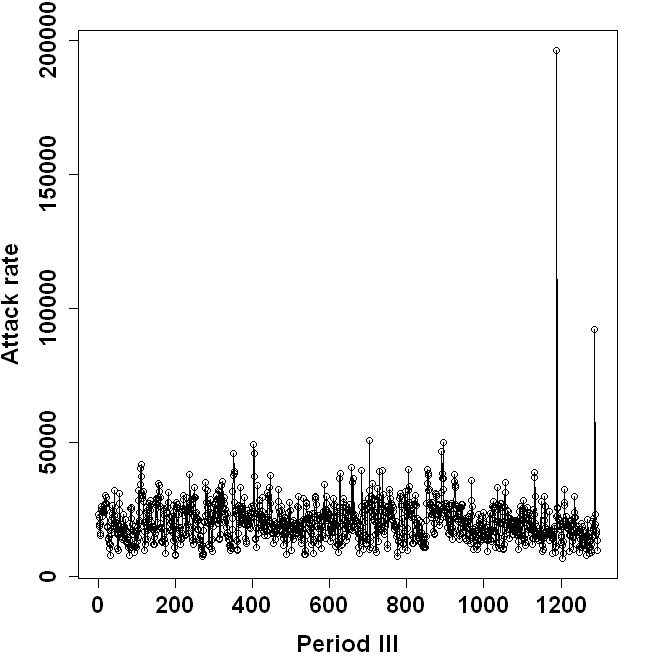} \label{ts-period-III}}
\subfigure[Period IV]{\includegraphics[width=.19\textwidth]{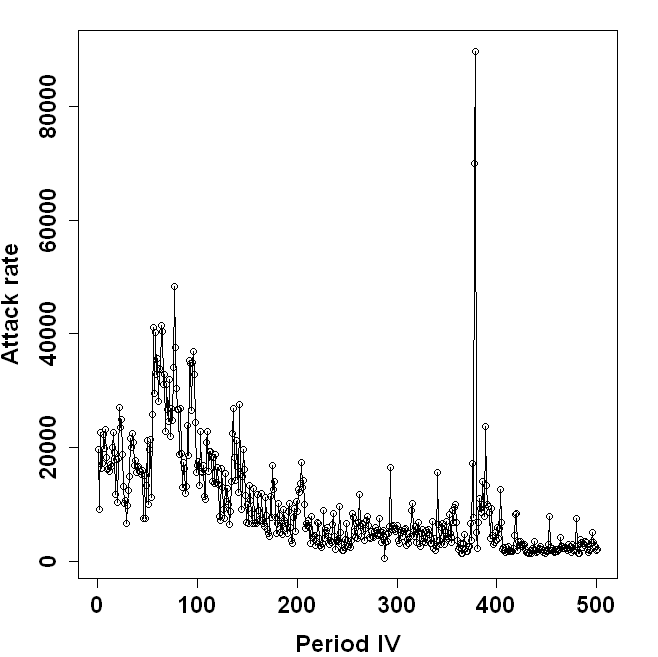}\label{ts-period-IV}}
\subfigure[Period V]{\includegraphics[width=.19\textwidth]{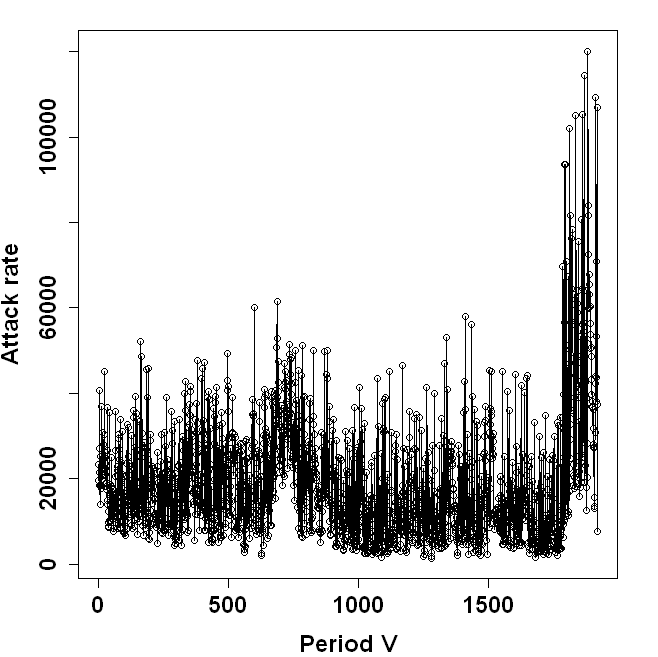} \label{ts-period-V}}
\caption{Time series plots of the network-level attack processes.
The $x$-axis indicates the relative time with respect to the start time for each period (unit: hour).
The $y$-axis indicates attack rate, namely the number of attacks (per hour) arriving at the honeypot.}
\label{fig:time-series-period-wise}
\end{figure*}

\section{Case Study}
\label{sec:case-study}

To demonstrate use of the framework, we conduct a case study by
applying it to analyze a dataset that was collected by a {\em low-interaction} honeypot. As mentioned above,
the framework can be equally applied to analyze {\em high-interaction} honeypot data.

\subsection{Data Pre-Processing}
\label{sec:data-description}

The dataset for our case study was collected by a honeypot,
which ran four popular low-interaction honeypot software programs:
Dionaea \cite{dionaea}, Mwcollector \cite{mwcollect}, Amun \cite{amun}, and Nepenthes \cite{nepenthes}.
The vulnerable services offered by all four honeypot programs are SMB, NetBIOS, HTTP,
MySQL and SSH, each of which is associated to a unique TCP port.
These are the production ports.
Each honeypot IP address was assigned to one of these programs and was completely isolated from the other honeypot IP addresses.
A single honeypot computer was assigned with multiple IP addresses to run multiple honeypot software programs.
A dedicated computer was used to collect the raw network traffic as {\tt pcap} files, which were timestamped at
the resolution of microsecond.
Table \ref{table:data-description} summarizes the dataset, which corresponds to 166 victim/honeypot IP addresses for five periods of time.
These periods are not strictly consecutive because of network/system maintenance etc.

\begin{table}[!hbtp]
\centering
{\footnotesize
\begin{tabular}{|c|c|p{.06\textwidth}|p{.08\textwidth}|}
\hline
Period & Dates  & Duration (days) & \# victim IPs\tabularnewline
\hline
I & 11/04/2010 - 12/21/2010 & 47  & 166\tabularnewline
\hline
II & 02/09/2011 - 02/27/2011 & 18  & 166\tabularnewline
\hline
III & 03/12/2011 - 05/06/2011 & 54  & 166\tabularnewline
\hline
IV & 05/09/2011 - 05/30/2011 & 21  & 166\tabularnewline
\hline
V & 06/22/2011 - 09/12/2011 & 80  & 166\tabularnewline
\hline
\end{tabular}
}
\caption{Data description}
\label{table:data-description}
\end{table}

In our pre-processing, we resolve the two issues described in the framework as follows.
First, we disregard the attacks against the non-production ports because such TCP connections are often dropped.
Note that the specific attacks against the production ports are dependent upon the
vulnerabilities emulated by the honeypot programs (e.g., Microsoft Windows Server Service Buffer Overflow
MS06040 and Workstation Service Vulnerability MS06070 for the SMB service).
Since low-interaction honeypots do not capture sufficient information for precisely recognizing the specific attacks,
we do not look into specific attack types.
Second, for flows that do not end with the FIN flag (indicating safe termination of TCP connection) or the RST flag (indicating unnatural termination
of TCP connection), we use the following two parameter values: 60 seconds for
the {\em flow timeout time}
and 300 seconds for the {\em flow lifetime}.

\subsection{Basic Statistical Analysis}

We consider the {\em per-hour} attack rate at three resolutions: the honeypot network,
individual victim IP address, and individual production port of each victim.
The choice of {\em per-hour} is natural, while noting that {\em per-day} attack rate is not appropriate
because each period is no more than 80 days.
Since the numbers of victim-level and port-level attack processes are much larger than the number of network-level attack processes,
different methods are used to represent their basic statistics.

\paragraph*{\bf Basic statistics of network-level attack processes}

For network-level attack processes, it is feasible and appropriate to plot the time series of the attack rate (per hour), namely
the total number of attacks against the honeypot network of 166 victims.
Figure \ref{fig:time-series-period-wise} plots the time series of attacks.
We make the following observations.
First, the five periods exhibit different attack patterns. For example, Periods I, II and V are relatively stationary.
Second, there are some extremely intense attacks during some hours in Periods III and IV.
The specific hour corresponding to the extreme value in Period III
is Apr 01, 2011, 12 Noon (US Eastern Time); the attacks are against the SSH services.
It is evident that the attacks are brute-forcing password.
The peak of attacks during Period IV occurs at
May 16, 2011, 3 AM (US Eastern Time). The intense attacks are against the HTTP service.
We find no information from the Internet whether or not there are worm/botnet outbreaks that correspond to the peaks.
Third, although the five plots exhibit some change-points, a formal statistical analysis (using the method reviewed in Section \ref{sec:preliminaries} for removing
spurious LRD)  shows that change-points exist
only in Period III, which correspond to the largest attack rate. This means that visual observations can be misleading
and rigorous statistical analysis is perhaps necessary.

Table \ref{table:period-centric-basic-statistics} describes the basic statistics of the network-level attack rate.
On average, the victim network is least intensively attacked during Period IV
because the average per-hour attack rate is about 9861, which is smaller than the average attack rate
during the other periods. The variances of attack rates are much larger than the corresponding mean attack rates,
which {\em hints} that these processes are not Poisson.
As we show via formal statistical analysis in Section \ref{sec:step-3}, these processes actually exhibit LRD instead.

\begin{table}[htbp]
{\footnotesize
\centering
\begin{tabular}{|r|r|r|r|r|r|}
\hline
Period & ${\sf MIN}$ & ${\sf Mean}$ & ${\sf Median}$ & ${\sf Variance}$ & ${\sf Max}$  \\
\hline
 I & 2572 & 30963.2 & 28263 & 401243263.2 & 151189 \tabularnewline
 \hline
 II & 5155 & 31576.8 & 29594 & 167872819.0 & 98527 \tabularnewline
\hline
 III & 6732 & 20382.3 &  19579 & 72436071.5 & 196210 \tabularnewline
 \hline
 IV & 637 & 9861.1 & 6528 & 93209085.3 & 89718 \tabularnewline
 \hline
 V  &  1417 & 18960.2 & 15248.5 & 205276388.4 & 120221 \tabularnewline
 \hline
 \end{tabular}
\caption{Basic statistics of network-level attack processes.}
\label{table:period-centric-basic-statistics}
}
\end{table}

\paragraph*{\bf Basic statistics of victim-level attack processes}

For victim-level attack processes, we consider the attack rate or the number of attacks (per hour) arriving at a victim.
Since there are 166 victims in each period, we cannot afford to plot time series of victim-level attack processes.

\begin{table}[htbp]
{\scriptsize
\centering
\begin{tabular}{|r|r|r|r|r|r|r|r|r|}
\hline
{\tiny Period}   &   \multicolumn{2}{|c|}{${\sf Mean}(\cdot)$} & \multicolumn{2}{|c|}{${\sf Median}(\cdot)$} %
& \multicolumn{2}{|c|}{${\sf Variance}(\cdot)$} & \multicolumn{2}{|c|}{${\sf MAX}(\cdot)$} \\ \cline{2-9}
 &   LB & UB & LB & UB & LB & UB & LB & UB \\
\hline
I  &  32.1 & 1810.4 & 8 & 1327 & 1589.9 & 3219758.8 & 247 & 14403 \\
\hline
II   & 49.8 & 1412.0 & 43 & 1112 & 1466.5 & 1553585.6 & 335 & 10995 \\
\hline
III  & 11.5 & 1513.5 & 3 & 1490 & 254.0 & 676860.7 & 125 & 5287 \\
\hline
IV  & 3.5 & 1663.4 & 1 & 1184 & 29.7 & 2808045.2 & 41 & 7793 \\
\hline
V   & 34.0 & 2228.8 & 8.5 & 1526.5 & 1225.6 & 4639659.1 & 274 & 12267 \\
\hline
\end{tabular}
\caption{Basic statistics of victim-level attack processes: attack rate (per hour).
For a specific period and a specific statistic $X\in\{{\sf Mean}, {\sf Median}, {\sf Variance},{\sf MAX}\}$,
LB (UB) stands for the lower-bound or minimum (upper-bound or maximum) of statistic $X$ among all the victims and all the hours.
In other words, the LB and UB values represent the minimum and maximum per-hour attack rate observed during an entire period and among all the victims.
}
\label{table:per-hour-attack-rate}
}
\end{table}

Table \ref{table:per-hour-attack-rate} summaries the observed lower-bound (minimum) and upper-bound (maximum) values of per-hour attack rate
for each statistic among the 166 victims.
By taking Period I as an example, we observe the following.
The average per-hour attack rate (among all the victims and among all the hours) is 32--1810 attacks per hour;
the median per-hour attack rate is 8--1327 attacks per hour;
the maximum number of attacks against a single victim can be up to 14403.
Boxplots of the four statistics, which are not included for the sake of saving space,
show that the five periods exhibit somewhat similar (homogeneous) statistical properties.
For example, each statistic has many outliers in each period.
By looking into all individual victim-level attack processes,
we find that among all the 830 victim-level attack processes (166 victims/period $\times$ 5 periods = 830 victims),
the variance of attack rate is at least 3.5 times greater than the mean attack rate corresponding to the same victim.
This fact --- the variance is much larger than the mean attack rate --- hints
that Poisson models may not be appropriate for describing victim-level attack processes.
This suggests us to conduct formal statistical tests, which will be presented in Section \ref{sec:step-3}.

\paragraph*{\bf Basic statistics of port-level attack processes}

For port-level attack processes,
Table \ref{table:basic-statistics-per-hour-attack-rate-without-considering-consecutive-attacks} summarizes the
lower-bound (minimum value) and upper-bound (maximum value) for each statistic.
By taking Period I as an example, we observe the following.
There can be no attacks against some production ports during some hours,
which explains why the {\sf Mean} per-hour attack rate can be 0.
On the other hand, a port (specifically, port 445 at Nov 6, 2010, 9 AM US Eastern time) can be attacked by 14363 attacks within one hour.
Like what is observed from the victim-level attack processes, we observe that the variance of attack rate is much larger than the mean attack rate.
This means that the port-level attack processes are not Poisson.
Indeed, as we will see in Section \ref{sec:step-5}, many port-level attack processes are actually heavy-tailed.

\begin{table}[!htbp]
{\scriptsize
\centering
\begin{tabular}{|r|r|r|r|r|r|r|r|r|}
\hline
{\tiny Period}   & \multicolumn{2}{|c|}{${\sf Mean}(\cdot)$} & \multicolumn{2}{|c|}{${\sf Median}(\cdot)$} %
& \multicolumn{2}{|c|}{${\sf Variance}(\cdot)$} & \multicolumn{2}{|c|}{${\sf MAX}(\cdot)$} \\ \cline{2-9}
 &  LB & UB & LB & UB & LB & UB & LB & UB \\
\hline
I  &   0 & 1740.7 & 0 & 1196 & 0 & 3249318.9 & 1 & 14363 \\
\hline
II &   0 & 1251.5 & 0 & 948 & 0 & 1545078.5 & 1 & 10992 \\
\hline
III &  0 & 1482.1 & 0 & 1458 & 0 & 661847.3 & 1 & 5275 \\
\hline
IV  &   0 & 1613.4 & 0 & 1142 & 0 & 2588396.6 & 1 & 6961 \\
\hline
V  &  0 & 2169.8 & 0 & 1448.5 & 0 & 4629744.3 & 1 & 12267 \\
\hline
\end{tabular}
\caption{Basic statistics of port-level attack processes: attack-rate (per hour).
As in Table \ref{table:per-hour-attack-rate}, LB and UB values represent the minimum and maximum per-hour attack
rate observed during an entire period and among all production ports of the victims.
}
\label{table:basic-statistics-per-hour-attack-rate-without-considering-consecutive-attacks}
}
\end{table}

\subsection{Identifying Statistical Properties of Attack Processes}
\label{sec:step-3}

We now characterize the statistical properties exhibited by network-level and victim-level attack processes.
In particular, we want to know they exhibit similar (if not exactly the same) or different properties.
In the above, we are already hinted that the attack processes are not Poisson. In what follows
we aim to pin down their properties.

\paragraph*{\bf Network-level attack processes exhibit LRD}

The hint that network-level attack processes are not Poisson suggests us to identify their properties.
It turns out that the network-level attack processes exhibit LRD as demonstrated by their Hurst parameters.
Table \ref{table:hurst-parameter-victim-network-centric} describes the six kinds of Hurst parameters corresponding to
the network-level attack processes. Although the Hurst parameters suggest that they all exhibit LRD,
a further analysis shows the LRD exhibited in Period III is spurious because it was caused by the non-stationarity of the process.
Therefore, 4 out of the 5 network-level attack processes exhibit (legitimate) LRD.

\begin{table}[hbt]
{\footnotesize
\centering
\begin{tabular}{|c|r|r|r|r|r|r|l|}
\hline
Period & ${\sf RS}$ & ${\sf AGV}$ & ${\sf Peng}$ & ${\sf Per}$ & ${\sf Box}$ & ${\sf Wave}$ & LRD? \\
\hline
 I & 0.80 & 0.95 & 0.88 & 1.03 & 1.00 & 0.75 & Yes \\
 \hline
 II & 0.74 & 0.59 & 0.86 & 0.75 & 0.97 & 0.84 & Yes\\
\hline
 III & 0.74 & 0.52 & 0.65 & 0.63 & 0.63 & 0.65 & No\\
 \hline
 IV & 1.05 & 0.97 & 0.95 & 1.07 & 0.97 & 1.22 & Yes\\
 \hline
 V &  0.74 & 0.78 & 0.74 & 1.03 & 0.80 & 0.80 & Yes\\
 \hline
 \end{tabular}
\caption{The estimated Hurst parameters for network-level attack processes.
The six estimation methods are reviewed in Appendix \ref{sec:hurst-parameter-estimation-methods}.
Note that a Hurst parameter value being negative or being greater than 1 means that either the estimation method is not suitable or the attack process is non-stationary.
}
\label{table:hurst-parameter-victim-network-centric}
}
\end{table}

\paragraph*{\bf Victim-level attack processes exhibit LRD}

For the 830 ($166$ victims/period $\times 5$ periods =830) victim-level attack processes,
we first rigorously show that they are not Poisson.
Assume that the attack inter-arrival times are independent and identically distributed
exponential random variables with distribution
$$F(x)=1-e^{-\lambda x},\, \lambda>0, x\ge 0.$$
To test the exponential distribution, we first estimate the unknown parameter $\lambda$ by the maximum likelihood method. Then, we compute the
Kolmogorov-Smirnov (KS), Cram\'{e}r-von Mises (CM), and Anderson-Darling (AD)
test statistics \cite{SW86,DS86} (cf. Appendix \ref{sec:goodness-of-fit-statistics} for a review) and compare them against the respective critical values.

\begin{table}[hbt]
{\footnotesize
\centering
\begin{tabular}{|c|r|r|r|r|r|r|r|r|r|r|r|r|}
\hline
Period    & \multicolumn{2}{|c|}{${\sf KS}$} & \multicolumn{2}{|c|}{${\sf CM}$} & \multicolumn{2}{|c|}{${\sf AD}$} \\ \cline{2-7}
 (days) & $\min$ & $\max$ & $\min$ & $\max$ & $\min$ & $\max$  \\
\hline
I  & 0.13 & 0.54 & 482.30 & 59543.87 & inf & inf \\
\hline
II & 0.06 & 0.50 & 47.08 & 20437.82 & 298.73 & inf \\
\hline
III & 0.06 & 0.65 & 163.71 & 51434.32 & 1103.70 & inf \\
\hline
IV  & 0.04 & 0.81 & 3.44 & 31376.27 & 22.83 & inf \\
\hline
V   & 0.08 & 0.65 & 323.39 & 214543.54 & inf & inf \\
\hline
CV  & \multicolumn{2}{|c|}{0.01} & \multicolumn{2}{|c|}{0.22} & \multicolumn{2}{|c|}{1.13} \\
\hline
\end{tabular}
\caption{Minimum values of the three test statistics for attack inter-arrival time (unit: second) corresponding to
the victim-level attack processes, where $\min$ and $\max$ represent the minimal and maximal minimum values among all
victim-level attack processes in a period, and ${\sf Inf}$ means the value is extremely large.
\label{table:minimum-statistics-inter-arrival-time}}
}
\end{table}

Table \ref{table:minimum-statistics-inter-arrival-time} reports the minimum test statistics,
where the critical values for the test statistics are based on significance level $.05$ and obtained from \cite{css81,cs2001}.
Since the values are far from the critical values, there is no evidence to support the exponential distribution hypothesis.
Because the minimum test statistics violate the
exponential distribution assumption already, greater test statistics must violate the
exponential distribution assumption as well.

\begin{figure}[hbt]
\centering
\subfigure[QQ-plot of inter-arrival time of victim-level attack process that exhibits the minimum KS, CM and AD value simultaneously]{
\includegraphics[width=.22\textwidth]{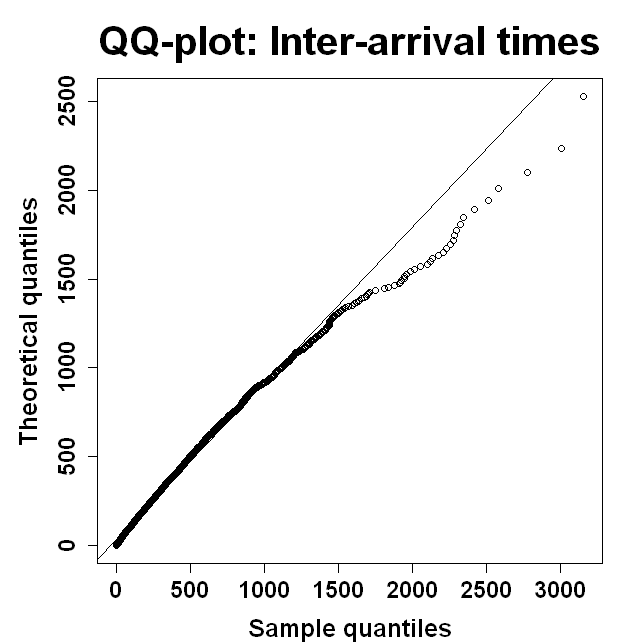}
\label{fig:qq-plot-qqqq}}
\subfigure[Boxplot of Hurst parameters of attack rate of the victim-level attack processes corresponding to the 5 periods]{
\includegraphics[width=.22\textwidth]{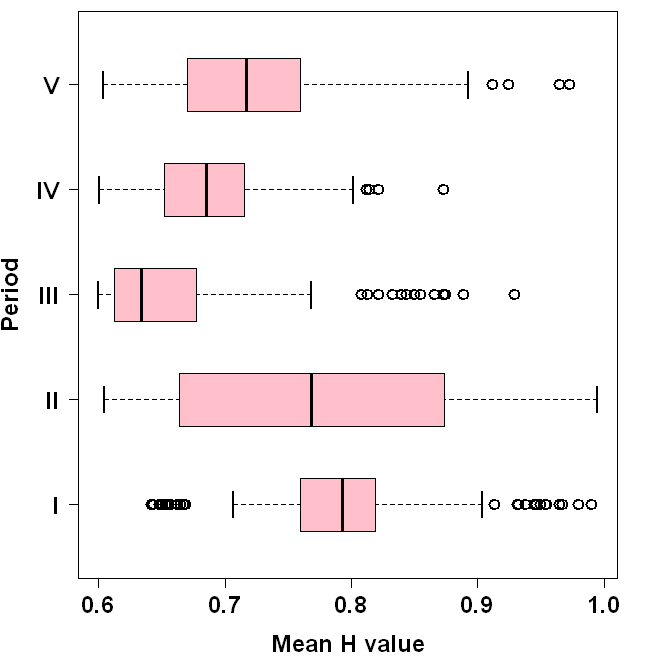}
\label{fig:Hboxplot-attack-rate}}
\caption{Victim-level attack processes are not Poisson but exhibit LRD}
\end{figure}

We also use QQ-plot to evaluate the goodness-of-fit of exponential distributions for the attack inter-arrival time of victim-level attack processes
that simultaneously exhibit the minimum test statistics in Table \ref{table:minimum-statistics-inter-arrival-time}.
This is the victim from Period IV with $H_{{\sf KS}}=0.04$, $H_{{\sf CM}}=3.44$ and $H_{{\sf AD}}=22.83$.
If the attack inter-arrival time corresponding to this particular victim does not exhibit the exponential distribution, we conclude that no attack inter-arrival time
in this dataset exhibits the exponential distribution. The QQ plot
is displayed in Figure \ref{fig:qq-plot-qqqq}. We observe a large deviation in the tails. Hence, exponential distribution
cannot be used as the distribution of attack inter-arrival times, meaning that all the victim-level attack processes are not Poisson.

\begin{table*}[!hbtp]
\centering
\begin{tabular}{|r|r|r|r|r|r|r|r|r|r|r|r|r|p{.12\textwidth}|p{.1\textwidth}|}
\hline
Period   & \multicolumn{2}{|c|}{${\sf RS}$} & \multicolumn{2}{|c|}{${\sf AGV}$} & \multicolumn{2}{|c|}{${\sf Peng}$} & \multicolumn{2}{|c|}{${\sf Per}$} &
\multicolumn{2}{|c|}{${\sf Box}$} & \multicolumn{2}{|c|}{${\sf Wave}$} & \# victims w/  & \# victims w/ \\ \cline{2-13}
   & $\min$ & $\max$ & $\min$ & $\max$ & $\min$ & $\max$ & $\min$ & $\max$ & $\min$ & $\max$ & $\min$ & $\max$ & $\bar{H}\in [.6,1]$ & LRD \\
\hline
I   & 0.53 & 1.01 & 0.46 & 0.98 & 0.66 & 1.14 & 0.73 & 1.39 & 0.55 & 1.15 & 0.40 & 0.96 & 163 & 159\\
\hline
II   & 0.49 & 0.94 & 0.40 & 0.98 & 0.56 & 1.37 & 0.53 & 1.69 & 0.33 & 1.32 & -0.55 & 1.33 & 130 & 116 \\
\hline
III   & 0.65 & 0.95 & 0.30 & 0.96 & 0.53 & 1.06 & 0.44 & 1.22 & 0.43 & 0.98 & 0.33 & 1.02 & 93 & 87 \\
\hline
IV   & 0.40 & 1.13 & 0.12 & 1.00 & 0.49 & 1.45 & 0.33 & 1.74 & 0.42 & 1.32 & -0.34 & 1.47 & 126 & 125 \\
\hline
V   & 0.52 & 1.01 & 0.14 & 0.99 & 0.45 & 1.22 & 0.47 & 1.43 & 0.57 & 1.30 & -0.16 & 1.18 & 158 & 89 \\
\hline
\end{tabular}
\caption{The estimated Hurst parameters for attack rate (per hour) of the victim-level attack processes.
The six estimation methods are reviewed in Appendix \ref{sec:hurst-parameter-estimation-methods}.
Note that a Hurst value being negative or being greater than 1 means that either the estimation method is not suitable or the process is non-stationary.
The column ``\# of victims w/ $\bar{H}\in [.6,1]$" represents the total number of victim-level attack processes
whose average Hurst parameters $\in [.6,1]$ (where average is among the six kinds of Hurst parameters),
which suggests the presence of LRD.
The column ``\# of victims w/ LRD" indicates the total number of victim-level attack processes
that exhibit LRD rather than spurious LRD. (The same notations will be used in the description of
Tables \ref{table:h-parameter-port-centric-attack-processes-production-port-only} and \ref{table:h-parameter-attack-rate-attacker-centric-production-ports-only}.)
\label{table:h-parameter-attack-rate-production-port-only}}
\end{table*}

Given that the victim-level attack processes are not Poisson, we suspect they might exhibit LRD as well.
Figure \ref{fig:Hboxplot-attack-rate} shows the boxplots of Hurst parameters of attack rate.
We observe that Periods I and II have relatively large Hurst parameters, suggesting stronger LRD.
Table \ref{table:h-parameter-attack-rate-production-port-only} summarizes the minimums and maximums of the estimated Hurst parameters
of attack rates. Consider Period I as an example, we observe that the attack processes corresponding to
163 (out of the 166) victims have average Hurst parameters falling into $[.6,1]$ and thus suggest LRD,
where the average is taken over the six kinds of Hurst parameters.
However, only 159 (out of the 163) victim-level attack processes exhibit legitimate LRD because the other
 4 (out of the 163) victim-level attack processes are actually spurious LRD (i.e., caused by the non-stationarity of the processes).
We also observe that in Period III, there are only 87 victim-level attack processes that exhibit LRD.
Overall, 70\% victim-level attack processes, or $159+116+87+125+89=576$ out of $166 \times 5=830$ attack processes,
exhibit LRD.

\paragraph*{\bf Port-level attack processes exhibit LRD}

Table \ref{table:h-parameter-port-centric-attack-processes-production-port-only}
summarizes the Hurst parameters of port-level attack processes.
We observe that there are respectively 316, 397, 399, 328, 406 port-level attack processes that exhibit LRD.
Since there are 5 production ports per victim and 166 victims, there are 830 port-level attack processes per period.
Since there are 5 periods of time, there are 4150 port-level attack processes in total (830 ports/period $\times$ 5 periods=4150 ports).
This means that $(316 + 397 + 399 + 328  406)/4150=44.5\%$ port-level attack processes exhibit LRD.

\begin{table*}[!hbtp]
\centering
\begin{tabular}{|r|r|r|r|r|r|r|r|r|r|r|r|r|r|r|r|}
\hline
Period   & \multicolumn{2}{|c|}{${\sf RS}$} & \multicolumn{2}{|c|}{${\sf AGV}$} & \multicolumn{2}{|c|}{${\sf Peng}$} & \multicolumn{2}{|c|}{${\sf Per}$} &
\multicolumn{2}{|c|}{${\sf Box}$} & \multicolumn{2}{|c|}{${\sf Wave}$} & total \# of & \# ports w/  & \#  ports w/ \\ \cline{2-13}
  & $\min$ & $\max$ & $\min$ & $\max$ & $\min$ & $\max$ & $\min$ & $\max$ & $\min$ & $\max$ & $\min$ & $\max$ & ports & $\bar{H}\in [.6,1]$ & LRD \\
\hline
I & 0.41 & 1.01 & -0.18 & 0.98 & -0.15 & 1.23 & 0.38 & 1.55 & 0.39 & 1.48 & -0.18 & 1.00 & 830 + 0 & 349 & 316\\
\hline
II& 0.23 & 1.50 & 0.04 & 0.97 & 0.18 & 1.51 & 0.32 & 1.68 & 0.26 & 1.45 & -0.60 & 1.38 & 829 + 1 & 419 & 397\\
\hline
III & 0.14 & 1.01 & -0.02 & 0.96 & 0.27 & 1.08 & 0.38 & 1.28 & 0.34 & 1.07 & 0.08 & 1.00 & 830 + 0 & 422 & 399\\
\hline
IV  & 0.25 & 1.17 & 0.05 & 1.00 & 0.24 & 1.57 & 0.18 & 1.70 & 0.29 & 1.50 & -1.10 & 1.72 & 828 + 2 & 339 & 328\\
\hline
V  & 0.43 & 1.14 & 0.12 & 0.99 & 0.42 & 1.40 & 0.45 & 1.52 & 0.40 & 1.41 & -1.07 & 1.43 & 830 + 0 & 528 & 406\\
\hline
\end{tabular}
\caption{The estimated Hurst parameters for port-level attack rate (per hour) of the port-level attack processes.
\label{table:h-parameter-port-centric-attack-processes-production-port-only}}
\end{table*}

\paragraph*{\bf Summary}

In summary, we observe that 80\% (4 out of 5) network-level attack processes exhibit LRD,
70\% victim-level attack processes exhibit LRD, and 44.5\% port-level attack processes exhibit LRD.
This means that defenders should expect that the burst of attacks will sustain,
and that cyber attack processes should be modeled using LRD-aware stochastic processes.

\subsection{Exploiting LRD to Predict Attack Rates}
\label{sec:step-4}

Assuming that the attacks arriving at honeypots are representative of, or related to, the attacks arriving at production networks (perhaps in some
non-trivial fashion that can be identified given sufficient data),
being able to predict the number of incoming attacks hours ahead of time can give the defenders sufficient early-warning time to prepare for the arrival of attacks.
Intuitively, the model that is good at prediction in this context should accommodate the LRD property.
This is confirmed by our study described below.

\paragraph*{\bf Prediction results for network-level attack processes}

In order to evaluate the accuracy of the prediction results, we use Algorithm \ref{evaluation-algorithm},
which is an instantiation of Algorithm \ref{prediction-algorithm} while considering prediction errors for evaluation purpose.
Let $\{X_1,\ldots,X_n\}$ be the time series of observed attack rates.
Algorithm \ref{evaluation-algorithm} uses portion of the observed attack rates $\{X_1,\ldots,X_n\}$ for fitting a prediction model
and compares the predicted attack rates to the observed attack rates for computing the prediction accuracy,
where ${\sf h}$ be an input parameter indicating the number of steps (i.e., hours) we will predict ahead of time,
and ${\sf p}$ be another input parameter indicating location of the prediction starting point.
In order to build reliable models, we set ${\sf p}= 50\%$ meaning that 50\% of the observed data is used as the training data for building models.

\begin{algorithm}[H]
\caption{Prediction Evaluation Algorithm}
\label{evaluation-algorithm}
INPUT: observed attack rates $\{X_1,\ldots,X_n\}$, ${\sf h}$ (hours ahead), ${\sf p}\in (0,1)$ indicates prediction start point\\
OUTPUT: prediction accuracy
\begin{algorithmic}[1]
\STATE {$t\gets \lfloor n*{\sf p}\rfloor$}
\WHILE{$t\leq n-{\sf h}$}
\STATE{Fit $\{X_1,\ldots,X_t\}$ to obtain an optimal model ${\sf M}_t$ as follows:
fit the data to 25 models FARIMA$(p,d,q)$ with varying parameters $p$ and $q$ (which uniquely determine parameter $d$)
 and select the best fitting model based on the AIC criterion \cite{CC2008}.}\COMMENT{The case of ARMA$(p,q)$ is similar.}
\STATE {Use ${\sf M}_t$ to predict $Y_{t+{\sf h}}$, the number of attacks that will arrive during the $(t+{\sf h})$th step}
\STATE {Compute prediction error $e_{t+{\sf h}}=X_{t+{\sf h}}-Y_{t+{\sf h}}$}
\STATE {$t\gets t+1$}
\ENDWHILE
\STATE{Compute ${\sf PMAD}$, ${\sf PMAD}'$, ${\sf OA}$, ${\sf UA}$ as defined in Section \ref{sec:measures}}
\RETURN{${\sf PMAD}$, ${\sf PMAD}'$, ${\sf OA}$, ${\sf UA}$}
\end{algorithmic}
\end{algorithm}

Now we report the prediction results, while comparing the LRD-aware FARIMA model and the LRD-less ARMA model.
Table \ref{table:prediction-period-centric-without-transformation} describes
 the prediction error of the network-level attack processes.
We observe the following.
First, for Periods I and II, both 1-hour ahead and 5-hour ahead FARIMA prediction errors are no greater than 22\%.
However, the 10-hour ahead FARIMA prediction is pretty bad.
This means that LRD-aware FARIMA can effectively predict the attack rate
even five hours ahead of time. This would give the defender enough early-warning time.

\begin{table}[btbp]
{\footnotesize
\begin{tabular}{|c|c|c|c|c|}
\hline
       & \multicolumn{2}{|c|}{${\sf PMAD}$} & \multicolumn{2}{|c|}{${\sf PMAD}'$} \tabularnewline
\hline
Period & ${\rm FARIMA}$ & ${\rm ARMA}$ & ${\rm FARIMA}$ & ${\rm ARMA}$ \tabularnewline
\hline
\multicolumn{4}{|c|}{1-hour ahead prediction (${\sf h}=1$, ${\sf p}=0.5$)} \tabularnewline
\hline
I & 0.179 & 0.446 & 0.173 & 0.157  \tabularnewline
\hline
II & 0.217 & 0.363 & 0.149 & 0.149 \tabularnewline
\hline
III & 0.298 & 0.273 & 0.305 & 0.312 \tabularnewline
\hline
IV & 0.548 & 0.526 & 0.126 &  0.106 \tabularnewline
\hline
V & 0.517 & 0.529 & 0.424 & 0.411 \tabularnewline
\hline
\multicolumn{4}{|c|}{5-hour ahead prediction (${\sf h}=5$, ${\sf p}=0.5$)} \tabularnewline
\hline
I & 0.206 & 0.556 & 0.292 & 0.314 \tabularnewline
\hline
II & 0.212 & 0.351 & 0.420 & 0.411 \tabularnewline
\hline
III & 0.297 & 0.272 & 0.246 & 0.250 \tabularnewline
\hline
IV & 0.847 & 0.838 & 0.226 & 0.207 \tabularnewline
\hline
V & 0.526 & 0.555 & 0.414 & 0.417 \tabularnewline
\hline
\multicolumn{4}{|c|}{10-hour ahead prediction (${\sf h}=10$, ${\sf p}=0.5$)} \tabularnewline
\hline
I & 0.869 & 0.801 & 0.314 & 0.281 \tabularnewline
\hline
II & 1.024 & 1.034 & 0.277 & 0.284 \tabularnewline
\hline
III & 1.00 & 1.002 & 0.202 & 0.201 \tabularnewline
\hline
IV & 0.648 & 0.627 & 0.282 & 0.490 \tabularnewline
\hline
V & 0.982 & 0.952 & 0.402 & 0.412\tabularnewline
\hline
\end{tabular}
\caption{Prediction error of network-level attack processes
using the LRD-aware FARIMA and the LRD-less ARMA, where prediction errors are defined in Section \ref{sec:preliminaries}.
${\sf p}=0.5$ means that we start predicting in the midpoint of each network-level attack process.}
\label{table:prediction-period-centric-without-transformation}
}
\end{table}

Second, Period III network-level attack process exhibits spurious LRD. However, both the LRD-aware FARIMA
and the LRD-less ARMA models can predict incoming attacks up to 5 hours ahead of time.
Indeed, the prediction error of FARIMA is slightly greater than
the prediction error of ARMA. This reiterates that if an attack process does not exhibit LRD,
it is better not to use LRD-aware prediction models;
if an attack process exhibits LRD, LRD-aware prediction models should be used.
This highlights the advantage of ``gray-box" prediction over ``black-box" prediction,
which demonstrates the principal utility of the statistical framework.

Third, although Period IV exhibits LRD, even its 1-hour ahead FARIMA
prediction is not good enough, with prediction error greater than 50\%.
While it is unclear what caused this effect, we note that the underestimation error ${\sf PMAD}'$
for 5-hour ahead prediction is still reasonable for Period IV
(22.6\% for FARIMA and 20.7\% for ARMA).
This means that if one is willing to over-provision defense resources to some extent,
then the prediction for Period IV is still useful.

\begin{table*}[htbp]
\centering
\begin{tabular}{|c|c|c|c|c|c|c|c|c|c|c|c|c|c|}
  \hline
    &total \# of victims &
\multicolumn{2}{|c|}{ \# of victims} &
\multicolumn{2}{|c|}{ \# of victims} &
\multicolumn{2}{|c|}{ \# of victims} &
\multicolumn{2}{|c|}{ \# of victims} &
\multicolumn{2}{|c|}{ \# of victims} &
\multicolumn{2}{|c|}{ \# of victims}  \\
  Period  &  ($(x_1,x_2)/(y)$) &
\multicolumn{2}{|c|}{w/ average } &
\multicolumn{2}{|c|}{w/ average} &
\multicolumn{2}{|c|}{w/ average} &
\multicolumn{2}{|c|}{w/ average} &
\multicolumn{2}{|c|}{w/ average} &
\multicolumn{2}{|c|}{w/ average}  \\
    &   &
\multicolumn{2}{|c|}{{\sf OA} $\geq 80\%$} &
\multicolumn{2}{|c|}{{\sf OA} $\geq 70\%$} &
\multicolumn{2}{|c|}{{\sf OA} $\geq 60\%$} &
\multicolumn{2}{|c|}{{\sf UA} $\geq 80\%$} &
\multicolumn{2}{|c|}{{\sf UA}  $\geq 70\%$} &
\multicolumn{2}{|c|}{{\sf UA} $\geq 60\%$}  \\
    \hline
&& {\tiny FARIMA} & {\tiny ARMA} & {\tiny FARIMA} & {\tiny ARMA} & {\tiny FARIMA} & {\tiny ARMA} & {\tiny FARIMA} & {\tiny ARMA} & {\tiny FARIMA} & {\tiny ARMA} & {\tiny FARIMA} & {\tiny ARMA} \\
\hline
\multirow{2}{*}{I}      &LRD: (152,152)/(159)  & 2  & 1  & 29 & 13 & 81 & 66 & 13 & 4 & 40 & 35 & 89 & 68 \\ \cline{2-14}
      &non-LRD: (7,7)/(7) & 0 & 0 & 4 &  4 & 6 & 6 & 1 & 4  & 7 &  6 & 7 &  7\\
\hline
\multirow{2}{*}{II}    &LRD: (109,109)/(116)  & 0  &0  & 3 &2 & 9 &8 &2 &1 &12 &6 &26 &15 \\ \cline{2-14}
      &non-LRD: (50,49)/(50) & 0 & 0 & 0 &  0 & 0 & 2 & 4 &  1 & 6 & 2 & 13 &  5 \\
\hline
\multirow{2}{*}{III}  & LRD: (82,82)/(87)  &0 & 0 &4 & 4 & 8 & 9 & 9 & 5 &23 & 19  &50 & 43  \\ \cline{2-14}
      &non-LRD: (79,79)/(79) & 0 & 0 & 0 &  0 & 0 & 0  & 0 & 0 & 10 & 7 & 31 & 24  \\
\hline
\multirow{2}{*}{IV}      &LRD: (118,118)/(125)  & 0 & 0 & 2 & 2  & 5 & 6 & 2 & 3 &4 & 6  & 11 &  14 \\ \cline{2-14}
      &non-LRD: (41,39)/(41) & 0 & 0 & 0 &  0 & 0 & 0 & 1 & 0 & 2 & 0 & 4 & 1 \\
\hline
\multirow{2}{*}{V}      &LRD: (73,73)/(89)  & 0 & 0 & 0 & 0 & 2 & 1 & 0 & 1 & 2 &  3 & 16 &  4  \\ \cline{2-14}
      &non-LRD:  (77,61)/(77)  & 0 &  0 & 0 & 0 & 1 & 1 & 0 & 0 & 1 & 0 & 24 &  15\\
\hline
\end{tabular}
\caption{Number of victim-level attack processes that can be predicted by the LRD-aware FARIMA model more accurately than the LRD-less ARMA model.
For the column ``total \# of victims ($(x_1,x_2)/(y)$),"
$y$ is the total number of victims that exhibited LRD or non-LRD, $x_1$ (or $x_2$) is total number of victims (out of the $y$ victims) for which the Maximum Likelihood Estimator (MLE)
used in the FARIMA (ARMA) algorithm converges (i.e., $y-x_1$ and $y-x_2$ victims cannot be predicted because the MLE does not converge).
The column ``\# of victims w/ average {\sf OA} (or {\sf UA}) $\geq z\%$" represents the average number of victims (out of the $x_1$ or $x_2$ victims that can be predicted),
for which the average prediction accuracy is at least $z\%$ in terms of overall-accuracy {\sf OA} (or underestimation-accuracy {\sf UA}),
where average is over all predictions.}
\label{table:victim-oriented-prediction-accurary}
\end{table*}

Fourth, Period V resists both prediction models in terms of both overall prediction error ${\sf PMAD}$ and underestimation error ${\sf PMAD}'$.
The fundamental cause of the effect is unknown at the moment, and is left for future studies.
Nevertheless, we suspect that Extreme Value Theory could be exploited to address this problem.

\paragraph*{\bf Prediction results for victim-level attack processes}

Since there are 166 victims per period, there are 830 victim-level attack processes for which we will do prediction.
Recall that 70\% victim-level attack processes exhibit LRD.
We use Table \ref{table:victim-oriented-prediction-accurary} to succinctly present the prediction results, which
are with respect to 10-hour ahead predictions during the last 100 hours of each time period.
We make the following observations.
First, the LRD-aware FARIMA model performs better than the LRD-less ARMA model.
For example, among the 152 (out of the 159) victim-level attack processes in Period I that exhibit LRD and are amenable to prediction (i.e., the Maximum Likelihood Estimator
actually converges; the Estimator does not converge for 159-152=7 LRD processes though), FARIMA can predict for 29 victim-level attack processes about their 10-hour ahead attack rates
with at least 70\% overall accuracy (${\sf OA}$), while ARMA can only predict for 13 victim-level attack processes at the same level of accuracy.
If the defender is willing to over-provision some resources and mainly cares about the
underestimation error (which could cause overlooking of attacks),
FARIMA can predict for 40 victim-level attack processes while ARMA can predict for 35.

Second, the victim-level attack processes in Period I exhibit LRD and render more to prediction when compared with the victim-level attacks processes in the other periods,
which also exhibit LRD. For non-LRD processes, neither FARIMA nor ARMA can provide good predictions.
This may be caused by the the non-stationary of the non-LRD processes.
We plan to investigate into these issues in the future.

\paragraph*{\bf Summary}

It is feasible to predict network-level attacks even 5 hours ahead of time.
For attack processes that exhibit LRD, LRD-aware models {\em can}  predict their attack rates better than LRD-less models do.
However, there are LRD processes that can resist the prediction of even LRD-aware models.
This hints that new prediction models are needed.

\begin{table*}[htbp]
\centering
{\begin{tabular}{ |c|r|r|r|r|r|r|r|r|r|r|r|r|r|r|}
  \hline
& total \# of& \# of victims w/ &  \multicolumn{5}{|c|}{\# of victims with} & total \# & Shape & \# of ports & \# of ports &  \\
  Period &  victims & sub-processes &  \multicolumn{5}{|c|}{certain \# of sub-processes } & of ports & mean & w/ shape & w/ shape & Standard  \\
& exhibiting & exhibiting &  \multicolumn{5}{|c|}{exhibiting heavy-tail} & exhibiting & value & value & value & deviation  \\ \cline{4-8}
& LRD&  heavy-tail &  1 & 2 & 3 & 4 & 5 & heavy-tail & & $\in (.5,1)$ & $\geq 1$ &  \\
    \hline
I & 159 & 56 & 50 & 6& 0&0 &0 &62 &.11 &1 & 0 &.11\\
\hline
II & 116 & 80 &78 & 11 &1 &0 &0 & 103&.40 & 50 &0 &  .22 \\
\hline
III & 87 & 47 &39 & 6& 2&0 &0 &57 &.22 & 2 &  0 &.18 \\
\hline
IV & 125 & 3 &  3 & 0& 0& 0& 0& 3&  .43 &1 & 0 &.35 \\
\hline
V & 89 &  32 & 29 & 1&2 & 0& 0& 37&.30 & 5 &  1 &.25 \\
\hline
\end{tabular}}
\caption{For victim-level attack processes exhibiting LRD, some port-level attack processes 
exhibit heavy-tails.
\label{table:port-exhibits-heavytails}}
\end{table*}

\subsection{Exploring (Non)Causes of LRD}
\label{sec:step-5}

Despite intensive studies in other settings, the fundamental cause of LRD is still mysterious.
One known possible cause of LRD is the superposition of heavy-tailed processes \cite{WillingerIEEEACMToN94,LelandInfocom91,WillingerIEEEACMToN97}.
Another candidate cause of LRD is that some attackers launch intense (consecutive) attacks (e.g., brute-forcing SSH passwords).
Now we examine the two candidate causes as described in the framework.

\paragraph*{\bf LRD exhibited by network-level attack processes is not caused by heavy-tailed victim-level attack processes}

We want to know whether or not the LRD
exhibited by the 4 network-level attack processes during Periods I, II, IV and V
are caused by the superposition of heavy-tailed victim-level attack processes.
That is, we want to know how many victim-level attack processes during each of the four periods are heavy-tailed.
We find that among the vector of $(166, 166, 166, 166)$ victim-level attack processes during Periods I, II, IV and V,
the vector of victim-level attack processes that exhibit heavy-tails is correspondingly $(101,0,24,31)$, 
by using the POT method that is reviewed in Appendix A-B.
This means that Period I is the only period during which majority of victim-level attack processes exhibit heavy-tails.
A few or even none processes in the three other periods exhibited heavy-tails.
This suggests that LRD exhibited by the network-level attack processes
does not have the same cause as what is believed for benign traffic \cite{Re07}.

\paragraph*{\bf LRD exhibited by victim-level attack processes is not caused by heavy-tailed port-level attack processes}
\label{sec:heavy-tail-cause}

Now we investigate whether or not the LRD exhibited by victim-level attack processes is caused by that the underlying port-level attack processes exhibit heavy-tails,
a property briefly reviewed in Appendix \ref{sec:heavy-tail-distributions}.
Table \ref{table:port-exhibits-heavytails} shows that only 8\% port-level attack processes,
or $56+80+47+3+32=218$ out of the $(159+116+87+125+89=576)$ victims $\times$ 5 ports/victim = 2880 port-level attack processes,
exhibit heavy-tails.
Moreover, only 29 (out of the 576) victim-level attack processes have 2 or 3 port-level attack processes that exhibit heavy-tails.
Further, there is only 1 port-level attack process that exhibits infinite mean because the shape value $\geq 1$,
and there are $1+50+2+1+5=59$ port-level attack processes that exhibit infinite variance because their shape values $\in (.5,1)$.
The above observations also hint that unlike in the setting of benign traffic \cite{Re07}, LRD exhibited by victim-level attack processes
is not caused by the superposition of heavy-tailed port-level attack processes.

\paragraph*{\bf LRD exhibited by victim-level attack processes is not caused by individual intense attacks}
\label{sec:consecutive-attack-cause}

Now we examine whether or not LRD is caused by the individual attackers that launch intense attacks.
For this purpose, we consider {\em attacker-level attack processes}, which model the attacks against each victim that are launched by {\em distinct} attackers.
In other words, we only consider the first attack launched by each attacker, while disregarding the subsequent attacks launched by the same attacker.

Table \ref{table:per-hour-attack-rate-without-considering-consecutive-attacks} describes the observed lower-bound and upper-bound
of the four statistics regarding the attacker-level processes, where the bounds are among all victims within a period of time.
By taking Period II as an example, we observe the following:
on average there are between 48 and 100 attackers against one individual victim within one hour,
and there can be up to 621 attackers against one individual victim within one hour.
Further, attacks in Periods III and IV exhibit different behaviors from the other three periods.
From the boxplots of the basic statistic, which are not presented for the sake of saving space, we observe that
the attackers' behaviors are actually very different in the 5 periods.
In particular, the attacker-level attack processes in Period II have many outliers in terms of the four statistics,
meaning that the attack rate during this period varies a lot.

\begin{table}[htbp]
{\footnotesize
\centering
\begin{tabular}{|r|r|r|r|r|r|r|r|r|}
\hline
Period     & \multicolumn{2}{|c|}{${\sf Mean}(\cdot)$} & \multicolumn{2}{|c|}{${\sf Median}(\cdot)$} %
& \multicolumn{2}{|c|}{${\sf Variance}(\cdot)$} & \multicolumn{2}{|c|}{${\sf MAX}(\cdot)$} \\ \cline{2-9}
  &   LB & UB & LB & UB & LB & UB & LB & UB \\
\hline
I   &  30.2 & 67.8 & 4 & 45 & 1498.1 & 4094.3 & 225 & 432 \\
\hline
II  & 48.6 & 100.8 & 42 & 93 & 1195.1 & 6298.3 & 306 & 621 \\
\hline
III   & 11.1 & 33.0 & 2 & 29 & 223.6 & 270.8 & 64 & 100 \\
\hline
IV   & 1.9 & 23.8 & 1 & 23 & 26.32 & 92.7 & 40 & 65 \\
\hline
V   & 33.4 & 127.9 & 8 & 105 & 1132.7 & 7465.2 & 266 & 605 \\
\hline
\end{tabular}
\caption{Basic statistics of attack rate of the attacker-level attack processes (per hour).}
\label{table:per-hour-attack-rate-without-considering-consecutive-attacks}
}
\end{table}

\begin{table*}[hbtp]
\centering
\begin{tabular}{|c|r|r|r|r|r|r|r|r|r|r|r|r|r|r|}
\hline
Period   & \multicolumn{2}{|c|}{${\sf RS}$} & \multicolumn{2}{|c|}{${\sf AGV}$} & \multicolumn{2}{|c|}{${\sf Peng}$} & \multicolumn{2}{|c|}{${\sf Per}$} &
\multicolumn{2}{|c|}{${\sf Box}$} & \multicolumn{2}{|c|}{${\sf Wave}$} & \# victims w/  & \# victims w/ \\ \cline{2-13}
   & $\min$ & $\max$ & $\min$ & $\max$ & $\min$ & $\max$ & $\min$ & $\max$ & $\min$ & $\max$ & $\min$ & $\max$ & $\bar{H}\in [.6,1]$ & LRD \\
\hline
I   & 0.593 & 0.977 & 0.851 & 0.958 & 0.896 & 1.111 & 1.174 & 1.334 & 0.942 & 1.185 & 0.582 & 0.843  & 153 & 153\\
\hline
II   & 0.570 & 0.883 & 0.616 & 0.950 & 0.689 & 1.070 & 0.710 & 1.152 & 0.663 & 1.242 & -0.360 & 0.728 & 92 & 77 \\
\hline
III   & 0.776 & 0.994 & 0.364 & 0.747 & 0.630 & 0.748 & 0.460 & 0.679 & 0.608 & 0.746 & 0.389 & 0.668 & 163 & 103\\
\hline
IV   & 0.657 & 0.920 & 0.273 & 0.955 & 0.690 & 0.872 & 0.559 & 1.206 & 0.612 & 0.952 & 0.288 & 1.004 & 166 & 165\\
\hline
V   & 0.495 & 0.758 & 0.563 & 0.727 & 0.499 & 0.806 & 0.898 & 1.114 & 0.660 & 0.977 & 0.567 & 0.931 & 166 & 77 \\
\hline
\end{tabular}
\caption{The estimated Hurst parameters of the attack rate of attacker-level attack processes (per hour).
\label{table:h-parameter-attack-rate-attacker-centric-production-ports-only}}
\end{table*}

In order to see whether or not the attacker-level attack processes still exhibit LRD,
we describe their Hurst parameters in Table \ref{table:h-parameter-attack-rate-attacker-centric-production-ports-only}.
Using Period I as an example, we observe that the attacker-level attack processes corresponding to 153 (out of the 166) victims 
suggest LRD because their average Hurst parameter $\in [.6,1]$,
where the average is taken over the six Hurst estimation methods. Moreover, none of the 153 attacker-level processes exhibit spurious LRD.
Using Period V as another example, we observe that all 166 attacker-level attack processes have average Hurst parameter $\in [.6,1]$,
but only 77 attacker-level attack processes exhibit LRD while the other 89 attacker-level attack processes exhibit
spurious LRD (caused by non-stationarity of the processes).
The above discussion suggests that LRD exhibited by victim-level attack processes is not caused by the intense (consecutive)
attacks launched by individual attackers, simply because most (or many) attacker-level attack processes also exhibit LRD.

\paragraph*{\bf Summary}

The LRD exhibited by stochastic cyber attack processes is neither necessarily caused by the superposition of heavy-tailed processes,
nor necessarily caused by the intense attacks launched by individual attackers.
While we ruled out these two candidate causes, it is an interesting and challenging future work
to precisely pin down the cause of LRD in this context.

\section{Discussion}
\label{sec:total-discussion}

In this section we discuss the limitation of the case study and the usefulness of the statistical framework.

\subsection{Limitation of the Case Study}
\label{sec:discussion}

The case study has three limitations that are imposed by the specific dataset.
First, the dataset, albeit over $47+18+54+21+80=220$ days in total (5 periods of time), only corresponds to 166 honeypot IP addresses.
We wish to have access to bigger datasets.
Still, this paper explores an important direction in cyber security research, especially
the feasibility of predicting incoming attacks.
Fortunately, the statistical framework
can be adopted by researchers to analyze their (bigger) datasets.

Second, the dataset is attack-agnostic in the sense that we know the ports/services the attackers attempt to attack,
but not the specific attacks because the data was collected using low-interaction honeypots.
Although this issue can be resolved by using high-interaction honeypots \cite{DBLP:journals/virology/NicometteKAH11},
there are legitimate concerns about high-interaction honeypots from a legal perspective.
Nevertheless, the framework is equally applicable to analyze high-interactive honeypot data.
For example, there might be researchers who have collected high-interaction honeypot data and are not allowed to share the data with others.
These researchers can adopt the framework to analyze their data at a finer resolution (e.g., the attack level that
an attack process can accommodate one or multiple families of attacks).

Third, the data is collected using honeypot rather than using production network.
For real-life adoption of the prediction capability presented in the paper,
attack traffic would be blended into the production traffic.
Whether or not the blended traffic also exhibits LRD is an interesting future study topic.
The main challenge again is the legal and privacy concerns in collecting such data.

\subsection{Usefulness of the Statistical Framework}

The usefulness of the statistical framework (or analysis methodology) can be seen from the following perspectives.
First, the framework has {\em descriptive} power because it aims to
study the advanced statistical properties exhibited by the cyber attack data that instantiates stochastic cyber attack process.
The advanced statistical properties are not known a prior. In order to obtain hints for the kinds of advanced
statistical properties that may be relevant, the framework starts at studying basic statistical properties exhibited by the data.
The hinted statistical properties are rigorously examined by using advanced statistical techniques,
which fall into the framework of Hypothesis Testing (e.g., whether LRD is exhibited or not is
tested based on the values of the Hurst parameters that are estimated using the rigorous statistical methods reviewed in Appendix \ref{sec:hurst-parameter-estimation-methods}).
Indeed, the framework guided us to identify the relevance of LRD in this aspect of cyber security, which is not known until now.

Second, the framework has {\em predictive} power because, as confirmed by the case study, it allows to exploit the newly identified
advanced statistical properties to predict the attack rate possibly hours ahead of time.
This kind of property-inspired ``gray-box" prediction, rather than ``black-box" prediction,
allows the defender to proactively provision defense resources.
Although the specific dataset used in our case study is collected by a low-interaction honeypot,
the concept of stochastic cyber attack process can equally describe the attacks that are observed at high-interaction honeypots.
Since high-interaction honeypots can collect more information about attacks, the framework can be equally applied to analyze the data with respect to specific attacks.
As a result, we can predict the arrival rate of specific attacks (i.e., attack-centric rather than computer/network-centric). Moreover, the framework in principle could
model and predict the emergence of new attacks (e.g., zero-day exploits), assuming the data exhibits LRD or other relevant
statistical properties that can be exploited for predicting the emergence of new attacks (e.g., the probability that a zero-day attack
will arrive at a honeypot, assuming that the data indeed contains new or zero-day attacks).
Although we do not have access to high-interaction
honeypot data, there would be researchers/practitioners who have access to such data.  This explains why we are automating our analysis methodology,
and will release the software package so that other researchers/practitioners can use our software package as is, or can enhance it to incorporate
more analysis methods to better serve their purposes.

Third, the framework can be adapted to describe attacks against production networks/computers, because
identifying statistical properties of the traffic would allow the defender to detect anomalies.  For example,
suppose the traffic during the past days does not exhibit LRD property (or any other relevant statistical property)
but the traffic today exhibits LRD property, then this hints possible attacks today.  Suppose further that a
firewall is installed to filter out the known attacks against the production networks/computers.  Then the change in statistical properties exhibited
by the traffic hints the presence of new (possibly zero-day) attacks against the production networks/computers.
These hints serve as clues for further forensics examinations.
Note that for further forensics examination of the actually attacks, we need the detailed information about the traffic.
Such information is not captured by low-interaction honeypots, but can be captured by high-interaction honeypots
and production defense systems.
This explains the limitation of our case study, although the limitation is not inherent to our framework (as it is imposed by the specific dataset
collected by low-interaction honeypot).

Fourth, consider the scenario that the honeypot IP addresses are randomly scattered into a production network
(rather than being allocated to a consecutive chunk of IP addresses).  This would be the ideal scenario for
deploying honeypots, because it can be hard for the attacker to figure out which IP addresses are the honeypot IP addresses.
This is true especially when the honeypot IP addresses are shuffled frequently and randomly and when the honeypot is a high-interaction one.
In this case, the attacks arrive at the honeypot IP addresses would be comparable to the attacks that arrive at the production IP addresses.
The attacks arrive at the honeypot IP addresses can be equally investigated using the framework presented in the paper. This means that we can predict the arrival
rate of the attacks that will come to the honeypot IP addresses possibly hours ahead of time.  This also means that we can expect the rate of attacks
that will arrive at the production IP addresses hours ahead of time.  When the predicted attack rate is high, the defender would need to provision
more defense resources to inspect the packets that target the production IP addresses (e.g., for deep packet inspection).
The prediction capability gives the defender early-warning that possibly intensive  attacks will arrive in the near future.
This kind of early-warning capability is desired for real-life defense.
Furthermore, the above discussion  equally applies to the case that known attacks have been filtered out by firewalls,
meaning that the prediction can be with respect to unknown (i.e., new or zero-day) attacks.

\section{Related Work}
\label{sec:related-work}

We discuss the related prior work from several perspectives.
In terms of analyzing honeypot-captured cyber attack data,
there have been at least two complementary approaches.
One approach is to visualize cyber attack data, such as using neural projection techniques to visualize the ports observed
in honeypot data \cite{HerreroIJNS12}. However, the widely used approach is statistical analysis.
Within this approach, existing studies mainly focused on the following aspects:
(i) analyzing attackers' probing activities \cite{PaxsonIEEETIFS2011};
(ii) grouping attacks (e.g., \cite{new.att.honeypot,honeypot.pca,cluster.cliques,iat.cliques,vis.att});
(iii) characterizing Internet threats \cite{DBLP:series/ais/AlataDDKKNPP06,DacierCoRR07} such as fitting
the attack inter-arrival time via a mixture of Pareto and Exponential distributions.
These studies are often based on flow-level processing of data, so do we in this paper.
In contrast to these studies, we systematically study the identification, exploitation and cause of statistical properties exhibited by honeypot data,
such as LRD that is shown to be exhibited by honeypot data for the first time in the present paper.
To our knowledge, our framework is the first formal statistical analysis of honeypot-captured cyber attack data.
In particular, our study of predicting cyber attacks (in terms of attack rate) would represent a significant step toward the ultimate goal of
quantitatively understanding/predicting cyber attacks.

In terms of using honeypots to improve defense, we note that honeypots have been used to help detect various attacks including DoS (denial-of-service) \cite{flow.dos},
worms \cite{DBLP:conf/raid/DagonQGLGLO04,flow.worm}, botnets \cite{flow.botnet,flow.botnet.WoNS,DBLP:journals/fgcs/PhamD11},
Internet-Messaging threats \cite{NDSS.IM.honeypot},
generating attack signatures \cite{Kreibich:2004:HCI:972374.972384,Portokalidis:2007:SZW:1224244.1224380}, and
detecting targeted attacks \cite{Anagnostakis:2005:DTA:1251398.1251407}.
These studies are important, but are orthogonal to the focus of the present paper.

In terms of the LRD phenomenon, we note that LRD was first observed in benign traffic
about two decades ago and there has been a large body of literature on this topic (e.g., \cite{WillingerIEEEACMToN94,LelandInfocom91,WillingerIEEEACMToN97,SamorodnitskyBook06}).
There have been studies on the effect of injecting abnormal events (which are not necessarily attacks) into benign traffics that exhibit LRD.
The injection of abnormal events may disrupt the LRD exhibited by the benign traffic (see, e.g., \cite{Nash:2001:SSN:2229240.2229729}).
There also have been studies on the effect of injecting attacks into benign traffics that exhibit LRD (in terms of number of bytes and number of packets).
The injection of attack events may not disrupt the LRD (i.e., the ``blended" traffics still exhibit LRD) \cite{AsSadhanSIIC08}.
In the setting of spams, the correlation co-efficient of inter-arrival time of spams that are sent by a group of spammers
may decrease slowly, which hints that the inter-arrival time may exhibit LRD \cite{Li_anempirical} --- although this was not formally statistically analyzed there.
In contrast to all the studies mentioned above, we investigate
LRD exhibited by {\em attack rate} via rigorous statistical methods: auto-correlation serves as a hint of possible LRD,
Hurst parameters serves as the first rigorous step of examining LRD, and non-stationarity analysis eliminates spurious LRDs.
To our knowledge, we are the first to report that LRD is  exhibited by honeypot-captured cyber attack data.

Putting data-driven analysis of cyber attacks into a broader context,
we note that there have been studies on characterizing blackhole-collected traffic data (e.g., \cite{PangIMC2004,BaileyIMC2010}) or one-way traffic in live networks
\cite{FontasIMC2012}. Still, there are no advanced statistical framework for analyzing such blackhole or one-way traffic data.
More specifically, these studies differ from ours in (i) honeypot-captured cyber attack data includes two-way communications,
whereas blackhole-collected data mainly corresponds to one-way communications;
(ii) we rigorously explore statistical properties such as LRD,
whereas their studies do not pursue such rigorous statistical analysis.
Nevertheless, it is possible that our analysis framework can be adapted to analyze blackhole data.

\section{Conclusion and Future Work}
\label{sec:conclusion}

We introduced the novel concept of stochastic cyber attack process, which offers a new perspective for studying cyber attacks and, in particular,
can be instantiated at multiple resolutions such as network-level, victim-level and port-level.
We then proposed a statistical framework that is centered on identifying, exploiting (for ``gray-box" prediction) and exploring (for cause analysis)
the advanced statistical properties of stochastic cyber attack processes.
In order to demonstrate use of the framework, we applied it to analyze some low-interaction honeypot  data.
The findings of the case study include: (i) majority of the attack processes exhibit LRD;
(ii) LRD-aware models can predict the attack rates (especially for network-level attack processes) even 5 hours ahead of time,
which would give the defender sufficient early-warning time.
The prediction power of the ``gray-box" prediction models, when compared with ``black-box" prediction models,
rewards the effort spent for analyzing the advanced statistical properties of stochastic cyber attack processes.

The present study introduces a range of interesting problems for future research.
First, we need to further improve the prediction accuracy, despite that the LRD-aware FARIMA model can
predict better than the LRD-less ARMA models.
For this purpose, we plan to study some advanced models that can accommodate high volatilities.
It is known in the literature that GARCH model may be able to accommodate high volatilities, which has some correlation  to LRD.
This hints that FARIMA+GARCH models may be able to fit and predict the
attack rates better. The other possible way to improve prediction is to incorporate the Extreme Value Theory into the
FARIMA process because the FARIMA process may not be able to capture the extremely large attack rate (i.e., the spikes).
Second, although our study only ruled out two candidate causes,
it is important to rigorously explain the fundamental cause of LRD as exhibited by honeypot-captured cyber attacks.
This is a difficult problem in general.
Third, the victim-level attack processes and network-level attack processes
exhibit similar phenomena (i.e., LRD). This hints a sort of {\em scale-invariance} that, if turns out to hold, would have extremely important
implications (for example) in achieving scalable analysis of cyber attacks.
Fourth, Wagener et al. \cite{WagenerSED11} recently introduced the concept of adaptive high-interaction honeypots to interact with the attacker
strategically, which is reminiscent of \cite{Cheswick92}.
It would be interesting to characterize the statistical properties of such new variants of honeypots.
Fifth, one reviewer points out the following interesting research problem: Can we use the information about the
prediction errors to directly adjust the prediction results? In order to answer this question, we need to
study the statistical properties of the prediction errors.

\section*{Acknowledgement}
This study was IRB-approved. We thank the anonymous reviewers for their comments that helped us improve the paper.
This work was supported in part by ARO Grant \#W911NF-13-1-0141 and AFOSR Grant \#FA9550-09-1-0165.
Any opinions, findings, and conclusions or recommendations expressed in this material are those of 
the author(s) and do not necessarily reflect the views of the funding agencies.

\appendices

\section{Review of Some Statistical Techniques}

\subsection{Methods for Estimating Hurst Parameters}
\label{sec:hurst-parameter-estimation-methods}

We used six popular methods (cf. \cite{Beran94} for details) for estimating the Hurst parameter,
which is a well-accepted practice \cite{TTW95,RRB2009}.

\smallskip

\noindent{1) {\sf RS} method}:
For a time series  $\{X_t, t\ge 1\}$,
      with partial sum $Y_t=\sum_{i=1}^t X_i$ and sample variance
       $S^2_t=\frac{1}{t}\sum_{i=1}^t X_i^2-\left(\frac{1}{t}\right)^2 Y_t^2$, the R/S statistic is defined as
       $$\frac{R}{S}(n)=\frac{1}{S_n}
       \left[\max_{0\le t\le n} \left(Y_t-\frac{t}{n}Y_n\right)-
       \min_{0\le t\le n}\left(Y_t-\frac{t}{n}Y_n\right)\right].$$
For  LRD  series, we have
 $$\E \left[\frac{R}{S}(n)\right]\sim C_H n^H,~~n\rightarrow \infty$$
where $C_H$ is a positive, finite constant independent of $n$.

\noindent{2) {\sf AGV} (aggregated variance) method}: Divide time series $\{X_t, t\ge 1\}$ into blocks of size $m$. The block average is
  $$X^{(m)}(k)=\frac{1}{m} \sum_{i=(k-1)m+1}^{km} X_i,\quad k=1,2\ldots.$$
Take the sample variance of $X^{(m)}(k)$ within each block, which is an estimator
 of ${\rm Var}(X^{(m)})$. For LRD series, we have $\beta=2H-2$ and
 $${\rm Var}\left(X^{(m)}\right)\sim c m^{-\beta},\quad m\rightarrow \infty,$$
where $c$ is a finite positive constant independent of $m$.

\noindent{3) {\sf Peng} method}: The series is broken up into blocks of size $m$. Compute partial sums $Y(i)$, $i=1,2\ldots,m$ within blocks. Fit a least-square
  line to the $Y(i)$'s and compute the sample variance of the residuals. This procedure is repeated for each of the blocks, and the resulting sample variances
  are averaged. The resulting number is proportional to $m^{2H}$ for LRD series.

\noindent{4) {\sf Per} (Periodogram) method}: One first calculates
$$I(\lambda)=\frac{1}{2\pi N}\left|\sum_{j=1}^N X_j e^{ij\lambda}\right|,$$
where $\lambda$ is the frequency, $N$ is the number of terms in the series, and $X_j$ is the data.
A LRD series should have a periodogram proportional to $\lambda^{1-2H}$ for $\lambda\approx 0$.
A regression of the logarithm of the periodogram on the logarithm of the frequency gives coefficient $1-2H$.

\noindent{5) {\sf Box} (Boxed Periodogram) method}: This method was developed to deal with the
problem that most points, which are used to estimate $H$, reside on the right-hand side of the graph.

\noindent{6) {\sf Wave} (Wavelet) method}:
Wavelets can be thought of as akin
to Fourier series but using waveforms other than sine
waves. The estimator used here fits a straight line to
a frequency spectrum derived using wavelets \cite{AV98}.

\subsection{Heavy-tail Distributions}
\label{sec:heavy-tail-distributions}

A random variable $X$ is said to belong to the Maximum Domain of Attraction (MDA) of the extreme value distribution $H_\xi$ if there exists constants $c_n\in \mathbb{R}_+$,
$d_n\in \mathbb{R}$ such that its distribution function $F$ that satisfies
$$\lim_{n\rightarrow \infty}  F^n(c_n x+d_n)=H_\xi(x).$$
 In statistics, $X$  is said to follow
a heavy-tailed distribution if $F\in {\rm MDA}(H_{\xi})$.
There are many methods for estimating parameter $\alpha$ \cite{Em1997,Re07}.
A widely-used method is called Point Over Threshold (POT).
Let $X_1,\ldots,X_n$ be independent and identically distributed random variables from $F\in {\rm MDA}(H_{\xi})$, then we may choose a high threshold $u$
such that
$$\lim_{u\rightarrow x_{F}}\sup_{0<x<x_F-u}|\bar F_u(x)-\bar G_{\xi,\beta(\mu)}(x)|=0,$$
where $x_F$ is the right end poind point of $X$, and
$$F_u(x)=P(X-u\le x|X>u), \quad x\ge 0,$$
and $ \bar G_{\xi,\beta(\mu)}=1-  G_{\xi,\beta(\mu)}$ is the survival function of generalized Pareto distribution (GPD)
$$\bar G_{\xi,\beta(\mu)}(x)=\left\{\begin{array}{cc}
  \left(1+\xi \dfrac{x}{\beta}\right)^{-1/\xi}, & \xi\ne 0 \\
  \exp\{-x/\beta\}, & \xi=0
\end{array}\right.
$$
where $x\in \mathbb{R}^{+}$ if $\xi\in\mathbb{R}^+$, and $x\in [0,-\beta/\xi]$ if $\xi\in \mathbb{R}^{-}$.
The POT method states that if $X_1,\ldots,X_n$ are heavy-tailed data, then $[X_i-u|X_i>u]$ follows a generalized Pareto distribution.

\subsection{Goodness-of-fit Test Statistics}
\label{sec:goodness-of-fit-statistics}

We use three popular goodness-of-fit test statistics:
Kolmogorov-Smirnov (KS), Cram\'{e}r-von Mises (CM), and Anderson-Darling (AD).
Let $X_1,\ldots,X_n$ be independent and identical random variables with distribution $F$. The empirical distribution $F_n$ is defined as
$$F_n(x)=\frac{1}{n} \sum_{i=1}^n {\rm I}(X_i\le x),$$
where ${\rm I}(X_i\le x)$ is the indicator function:
$${\rm I}(X_i\le x)=\left\{\begin{array}{cc}
  1, & X_i\le x, \\
  0, & o/w.
\end{array}\right.
$$
The KS test statistic is defined as
$${\rm KS}=\sqrt{n}\sup_{x}\left|F_n(x)-F(x)\right|.$$
The CM test statistic is defined as
$${\rm CM}=n\int (F_n(x)-F(x))^2 d F(x).$$
The AD test statistic is defined as
$${\rm AD}=n\int (F_n(x)-F(x))^2 w(x)d F(x),$$
where $w(x)=[F(x)(1-F(x))]^{-1}$.

\begin{wrapfigure}{l}{0.15\textwidth}
\centering
\includegraphics[height=0.12\textwidth,width=0.12\textwidth]{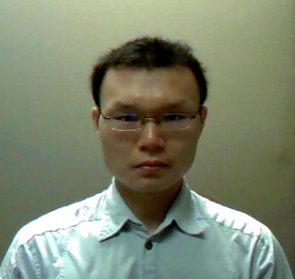}
\end{wrapfigure}
Zhenxin Zhan is a PhD candidate in the Department of Computer Science, University of Texas at San Antonio.
He received M.S. degree in Computer Science from the Huazhong University of Science and Technology, China, in 2008.
His primary research interests are in cyber attack analysis and detection.

\begin{wrapfigure}{l}{0.15\textwidth}
\centering
\includegraphics[height=0.12\textwidth,width=0.12\textwidth]{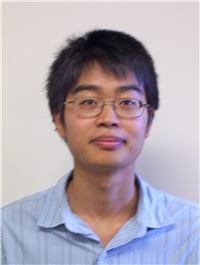}
\end{wrapfigure}
Maochao Xu received his PH.D. in Statistics from Portland State University in 2010. He is an Assistant Professor of Mathematics at the Illinois State University. 
His research interests include Applied Statistics, Extreme value theory,  Cyber security, and Risk analysis in actuary and insurance. 
He currently serves as an associate editor for Communications in Statistics.

\begin{wrapfigure}{l}{0.15\textwidth}
\centering
\includegraphics[height=0.12\textwidth,width=0.12\textwidth]{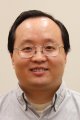}
\end{wrapfigure}
Shouhuai Xu is an Associate Professor in the Department of
Computer Science, University of Texas at San Antonio.
His research interests include cryptography and
cybersecurity modeling \& analysis.
He earned his PhD in Computer Science from Fudan University, China.
More information about his research can be found at
\url{www.cs.utsa.edu/~shxu}.

\end{document}